\documentclass[12pt,preprint]{aastex}
\usepackage{amsmath,bm}
\usepackage{multirow}
\usepackage{enumerate}
\usepackage{graphicx}
\usepackage{color}
\usepackage[T1]{fontenc}
\usepackage{url}
\usepackage{hyperref}
\newcount\doepsf
\newcommand{\be}{\begin{equation}}
\newcommand{\ee}{\end{equation}}
\newcommand{\bea}{\begin{eqnarray}}
\newcommand{\eea}{\end{eqnarray}}
\newcommand{\bean}{\begin{eqnarray*}}
\newcommand{\eean}{\end{eqnarray*}}

 \definecolor{DarkGreen}{rgb}{0.0,0.45,0.0}     
 \definecolor{DarkMagenta}{rgb}{0.45,0.0,0.45}  

\begin{document}

\title{Fast magnetic reconnection in the solar chromosphere mediated by the plasmoid instability}

\author{Lei Ni$^{1,2}$,
        Bernhard Kliem$^{1,3}$,
        Jun Lin$^{1}$,
    and Ning Wu$^{4}$}

\affil{$^1$Yunnan Observatories, Chinese Academy of Sciences,
                 Kunming 650011, China}
\affil{$^2$Key Laboratory of Solar Activity, National
           Astronomical Observatories, Chinese Academy of Sciences,
           Beijing 100012, China}
\affil{$^3$Institute of Physics and Astronomy, University of Potsdam,
                 Potsdam 14476, Germany}
\affil{$^4$School of Tourism and Geography, Yunnan Normal University,
           Kunming 650031, China}


\shorttitle{Fast reconnection in the solar chromosphere}  
\shortauthors{Ni et al.}

\email{leini@ynao.ac.cn}

\slugcomment{ChromPlasmoids; v.\ \today}

\begin{abstract}

\noindent  Magnetic reconnection in the partially ionized solar chromosphere is studied in 2.5-dimensional magnetohydrodynamic simulations including radiative cooling and ambipolar diffusion. A Harris current sheet with and without a guide field is considered. Characteristic values of the parameters in the middle chromosphere imply a high magnetic Reynolds number of $\sim10^{6}\mbox{--}10^7$ in the present simulations. Fast magnetic reconnection then develops as a consequence of the plasmoid instability without the need to invoke anomalous resistivity enhancements. Multiple levels of the instability are followed as it cascades to smaller scales, which approach the ion inertial length. The reconnection rate, normalized to the asymptotic values of magnetic field and Alfv\'en velocity in the inflow region, reaches values in the range $\sim0.01\mbox{--}0.03$ throughout the cascading plasmoid formation and for zero as well as for strong guide field. The out-flow velocity reaches $\approx40$~km\,s$^{-1}$.  Slow-mode shocks extend from the $X$-points, heating the plasmoids up to $\sim 8\times10^4$~K. In the case of zero guide field, the inclusion of ambipolar diffusion and radiative cooling both cause a rapid thinning of the current sheet (down to $\sim30$~m) and early formation of secondary islands. Both of these processes have very little effect on the plasmoid instability for a strong guide field. The reconnection rates, temperature enhancements, and upward out-flow velocities from the vertical current sheet correspond well to their characteristic values in chromospheric jets.

\end{abstract}   


\section{Introduction}
\label{s:introduction}
It is well known that magnetic reconnection plays an important role in the dynamic solar corona, where it is a key process in eruptions (flares, filament eruptions, coronal mass ejections), jets, and other phenomena.  In the lower solar atmosphere, magnetic reconnection is also very common in different forms of activity, such as microflares (e.g., Brosius \& Holman 2009; Gontikakis, Winebarger, \& Patsourakos 2013), the flux cancellation process in the photosphere (e.g., Zwaan1987; van Ballegooijen \& Martens 1989; Cameron, V{\"o}gler, \& Sch{\"u}ssler2007), chromospheric jets (e.g., Chae, Moon, \& Park 2003; Shibata et al. 2007;  Liu et al. 2009; Shen et al. 2011; Shen et al. 2012; Morton 2012; Bharti, Hirzberger, \& Solanki 2013), and temperature enhancements in Ellerman bombs (e.g., Zachariadis, Alissandrakis, \& Banos1987; Denker1997; Dara et al.1997; Qiu et al.2000; Georgoulis et al. 2002; Fang et al. 2006). Magnetic reconnection probably plays an important role as the energy source for the heating of the chromosphere and corona (e.g., Parker 1972; Sturrock 1999; Klimchuk 2006). The source of the mass and energy in the solar wind may also be related to magnetic reconnection in the lower solar atmosphere.

Compared to the solar corona, the mass density and plasma $\beta$ in the chromosphere and photosphere are higher,  and the temperature is much lower, mostly below $10^4$~K, except in the upper chromosphere. Therefore, the plasma in this height range is only partially ionized with an ionization degree of about $10^{-4}\mbox{--}10^{-1}$ \cite[e.g.,][]{1981ApJS...45..635V, 1986A&A...154..231P, 2008SoPh..251..589K}. The magnetic diffusivity is of order $\eta\sim(10^3\mbox{--}10^4)$~m$^2$\,s$^{-1}$ in the range from the upper photosphere to the middle chromosphere, so that the Lundquist number (magnetic Reynolds number) for a length-scale $L\sim1$~Mm and Alfv\'en velocity $v_\mathrm{A}\sim(10\mbox{--}100)$~km\,s$^{-1}$ is of order $S=Lv_\mathrm{A}/\eta\sim10^{6}\mbox{--}10^8$. The classical Sweet-Parker model for a current sheet of this length yields a reconnection rate $\gamma_\mathrm{SP}\sim S^{-1/2}\sim10^{-4}\mbox{--}10^{-3}$, far smaller than the reconnection rates inferred for events of chromospheric activity. For example, \citet{2011ApJ...731...43N} estimated $\gamma\sim0.02\mbox{--}0.1$ for chromospheric anemone jets from their life time and the estimated Alfv\'en crossing time. The same estimate for type-II spicules, which have life times of 10--150~s, sizes of $\sim\!200$~km, and ambient Alfv\'en velocities based on the total density of $\sim\!100$~km\,s$^{-1}$ \citep{2007PASJ...59S.655D, 2011ApJ...731L..18M}, yields $\gamma\sim0.01\mbox{--}0.2$. Such and even larger disparity is typically found for weakly ionized astrophysical plasmas like, for example, the interstellar medium and protostellar and protoplanetary disks (\citealt{2011PhPl...18k1211Z} and references therein). 

Many numerical simulations applied to phenomena of solar activity employ `anomalous' resistivity enhancements to obtain fast reconnection. However, the exact forms of anomalous resistivity used in magnetohydrodynamic (MHD) models are not directly deducible from the kinetic theory and simulations of real physical systems, and therefore some simplifying assumptions are still required for the model of anomalous resistivity (see, e.g., \citealt{2012ApJ...758...20N} and references therein).

Steady-state reconnection rates in weakly ionized plasma can be significantly enhanced above the Sweet-Parker value by radiative cooling and by ambipolar diffusion; however, both are efficient only if the guide field is weak. Radiative cooling reduces the pressure in the current sheet, thus allowing the sheet to thin and, correspondingly, the current density to rise. This implies higher Lorentz forces accelerating the reconnection outflow, resulting in an enhancement over the Sweet-Parker rate of order $A^{1/2}$, where $A$ is the compression factor of the sheet due to the cooling \citep{1995ApJ...449..777D, 2011PhPl...18d2105U}. In the incompressible strong guide field case, the cooling leads to a weaker enhancement of the reconnection rate through the higher resistivity, but in a chromospheric environment this can hardly reach an order of magnitude. Ambipolar diffusion in the current sheet decouples the ions and neutrals, so that only the ion pressure is available to balance the Lorentz force, which results in a strong current sheet thinning for weak ionization, allowing rapid flux annihilation \citep{1994ApJ...427L..91B}. The enhancement over the Sweet-Parker reconnection rate can be very large in the special case of antiparallel field \citep{2003ApJ...583..229H}, however, already a tiny guide field component suppresses this fast reconnection regime \citep{2003ApJ...590..291H}.

In partially ionized plasma, fast reconnection also results from the recombination effect under non-equilibrium ionization conditions. The numerical results from a two fluid model in the papers by \citet{2012ApJ...760..109L, 2013PhPl...20f1202L} indicate that the reconnection rate can been increased to above 0.05 solely due to the decoupling of neutrals from ions. Therefore, when the neutral-ion collisional mean free path exceeds the current sheet width, the two-fluid model including the recombination effect should be applied in studying the reconnection process.

Reconnection also becomes fast when the current sheet thins down to the ion inertial scale $d_\mathrm{i}$ where the Hall term is relevant \citep{1994GeoRL..21...73M, 2011ApJ...739...72M}. However, a dynamic regime of reconnection which involves plasmoids (magnetic islands) realizes high reconnection rates already at current sheet widths typically several orders of magnitude larger than $d_\mathrm{i}$; this is now commonly referred to as the plasmoid instability \citep{2007PhPl...14j0703L, 2009PhRvL.103f5004D, 2009PhPl...16k2102B}. The motion of the plasmoids in the current sheet is largely constrained by ideal MHD \citep{1977PhFl...20...72F}; hence, it is fast and largely independent of the resistivity, forming a highly efficient internal engine of fast reconnection. The instability cascades to smaller scales such that current sheet sections between newly formed islands break up as well; the resulting high current densities in the increasingly thin current sheet fragments also facilitate a high reconnection rate. Moreover, this mechanism operates for current sheets with and without a guide field. Its high relevance for reconnection in the solar corona, especially for flares, has recently been demonstrated in a number of simulation studies (e.g., Murphy et al. 2012; Mei et al. 2012; Guo et al. 2013). The plasmoid instability has recently been found to occur also in simulations of reconnection in weakly ionized plasma \citep*{2012ApJ...760..109L, 2013PhPl...20f1202L}. Plasmoids indicating the action of the instability were detected in laboratory experiments \cite[e.g.,][]{2012PhRvL.108u5001D}, in the solar corona in the course of eruptions (e.g., Lin, Cranmer, \& Farrugia 2008; Savage et al. 2010; Milligan et al. 2010; Takasao et al. 2012; Liu 2013), as well as in chromospheric jets \citep{2007Sci...318.1591S, 2012ApJ...759...33S}. 


The plasmoid instability occurs only if the Lundquist number and the aspect ratio $L/\delta$ of the current sheet are sufficiently large, where $\delta$ is the current sheet width. Both are related through the Sweet-Parker scaling for the current sheet width, $\delta_\mathrm{SP}\sim S^{-1/2}L$. For fully ionized plasma, the critical Lundquist number, $S_\mathrm{cr}$, typically lies in the range of several $10^3$ to several $10^4$, decreasing for increasing plasma $\beta$ in the inflow region (e.g., Bhattacharjee et al. 2009; Huang \& Bhattacharjee 2010; Ni et al. 2010; Ni et al. 2012;  Ni et al. 2013). The minimum aspect ratio correspondingly is of order $\sim10^2$. Fast reconnection at a rate $\gamma\sim S_\mathrm{cr}^{-1/2}$ was found for $S>S_\mathrm{cr}$. \citet{2012ApJ...760..109L, 2013PhPl...20f1202L} find the onset of the plasmoid instability in partially ionized plasma in basic agreement with this value.

Since the characteristic Lundquist number estimated for chromospheric reconnection events is much higher than the critical Lundquist number for onset of the plasmoid instability, and since this process yields high reconnection rates for weak and strong guide fields, we focus on the plasmoid instability in chromospheric reconnection in the present investigation, considering both the regimes of strong and of vanishing guide field. Parameters characteristic of the middle solar chromosphere and a low plasma $\beta$ in the inflow region are considered. The energetically and dynamically important processes of radiative cooling and ambipolar diffusion are included, and their effect on the reconnection rate is compared with the effect of the cascading plasmoid instability. The formation of slow-mode shocks associated with the X-points between the plasmoids is demonstrated. Finally, we compare the obtained reconnection rates, upward reconnection outflow velocities, and temperature enhancements with the values typically seen in or inferred for events of chromospheric activity.

The model and numerical method are described in the following section. In Section~3, we present our numerical results and compare them with observations of chromospheric activity phenomena. A  summary and discussion are given in Section~4.

\section{Model and Numerical Method}
\label{s:model}
We only consider the hydrogen gas, so that the plasma is composed of neutral hydrogen atoms, ions (protons) and electrons. In general this system is described by the three-fluid MHD equations. However, a single-fluid approximation adding up all three equations can be used if the collisional coupling between ions and neutrals is strong, which is valid on length-scales exceeding the neutral-ion collisional mean free path $\lambda_\mathrm{ni}$. The relevant length-scale here is the current sheet width $\delta$. We will find below that parameter values characteristic of the middle chromosphere yield critical current sheet widths for onset of the plasmoid instability several orders of magnitude above $\lambda_\mathrm{ni}$, indicating that the onset and initial evolution of the instability in this part of the solar atmosphere can generally be described using the one-fluid equations. Since the instability tends to cascade to small scales comparable to the ion inertial length, which can be of similar magnitude as $\lambda_\mathrm{ni}$ or smaller (depending on the ion density), it is clear that a complete description of plasmoid-mediated reconnection in the chromosphere will require a multi-fluid treatment (such as recently developed by \citealt{2012ApJ...760..109L, 2013PhPl...20f1202L}). However, the present investigation demonstrates that the plasmoid instability occurs and yields high reconnection rates already within the validity range of the one-fluid approximation adopted here. Since the neutrals are coupled to the ions in this approximation, the kinematic behaviour of the plasma bears a considerable degree of similarity to the fully ionized case, but radiative losses and ambipolar diffusion can nevertheless influence the dynamics strongly.

Additionally, we approximate the gas pressure tensor as a scalar, assume isothermal conditions (for simplicity and in the absence of detailed information to the opposite), $T_\mathrm{i}=T_\mathrm{n}=T_\mathrm{e}$, and neglect heat conduction. The latter is based on the fact that heat conduction is energetically less important than radiative cooling for the high densities and low temperatures characteristic of the middle chromosphere \citep{1992A&A...253..557L}. Additionally, heat conduction into and out of plasmoids is largely across the magnetic field, especially in a two-dimensional description, thus, it tends to be small (see Section~\ref{s:discussion} for a detailed discussion). This results in the set of basic equations \cite[see, e.g.,][]{2012ApJ...747...87K}

\begin{eqnarray}
 \partial_t \rho &=& -\nabla \cdot (\rho \bm{v})                                                                \\ 
  \partial_t \bm{B} &=& \nabla \times (\bm{v} \times \bm{B}-\eta\nabla \times \bm{B}+\bm{E}_\mathrm{AD})\label{e:induction}\\
  \partial_t (\rho \bm{v}) &=& -\nabla \cdot \left[\rho \bm{v}\bm{v}
                              +\left(p+\frac {1}{2\mu_0} \vert \bm{B} \vert^2\right)\mbox{\bfseries\sffamily I} \right]  \nonumber \\
                              & &+\nabla \cdot \left[\frac{1}{\mu_0} \bm{B} \bm{B} \right] + \rho \bm{g} \\
 \partial_t e &=& - \nabla \cdot \left[ \left(e+p+\frac {1}{2\mu_0 }\vert \bm{B} \vert^2\right)\bm{v} \right] \nonumber\\
                           & &+\nabla \cdot \left[\frac {1}{\mu_0} (\bm{v} \cdot \bm{B})\bm{B}\right]             \nonumber \\
              & & + \nabla \cdot \left[ \frac{\eta}{\mu_0} \textbf {B} \times (\nabla \times \bm{B}) \right] \nonumber\\
                                       & &-\nabla \cdot \left[\frac{1}{\mu_0}\bm{B} \times \bm{E}_\mathrm{AD}\right] \nonumber\\
                                       & &+\rho \bm{g} \cdot \bm{v}+\mathcal{L}_\mathrm{rad}+\mathcal{H}   \\
   e &=& \frac{p}{\Gamma_0-1}+\frac{1}{2}\rho \vert  \bm{v} \vert^2+\frac{1}{2\mu_0}\vert \bm{B} \vert^2          \\
   p &=& \frac{(1+Y_\mathrm{i}) \rho}{m_\mathrm{i}} k_\mathrm{B}T .
\end{eqnarray}
As usual, $\rho$ is the plasma mass density, $\bm{v}$ is the centre of mass velocity, $e$ is the total energy density, $\bm{B}$ is the magnetic field, $\eta$ is the magnetic diffusivity, and $p$ is plasma thermal pressure, $\bm{g}=-273.9~\mbox{m\,s}^{-2}~\bm{e}_y$ is the gravitational acceleration of the Sun. The ambipolar electric field is given by $\bm{E}_\mathrm{AD} = \mu_0^{-1}\eta_\mathrm{AD}[(\nabla \times \bm{B}) \times \bm{B}]\times \bm{B}$, where $\eta_\mathrm{AD}$ is the ambipolar diffusion coefficient.  $\mathcal{L}_\mathrm{rad}$ is the radiative cooling function and $\mathcal{H}$ is the heating function. $Y_\mathrm{i} = n_\mathrm{e}/n_\mathrm{H}$ is the ionization degree of the plasma.
In the above equations, the following definitions are used:
\begin{eqnarray}
 \rho &=& \rho_\mathrm{n}+\rho_\mathrm{i}+\rho_\mathrm{e} \simeq \rho_\mathrm{H},                              \\
 m_\mathrm{i} &\simeq& m_\mathrm{n},                                                                           \\
 \bm{v} &=& \frac{\rho_\mathrm{n} \bm{v}_\mathrm{n} +\rho_\mathrm{i} \bm{v}_\mathrm{i} + \rho_\mathrm{e} \bm{v}_\mathrm{e}}{\rho}, \\ 
 p &=&  n_\mathrm{n}k_\mathrm{B}T_\mathrm{n}+n_\mathrm{i}k_\mathrm{B}T_\mathrm{i}+n_\mathrm{e}k_\mathrm{B}T_\mathrm{e} \nonumber \\
    &=&  (n_\mathrm{n}+n_\mathrm{i}+n_\mathrm{e})k_\mathrm{B}T                                                \nonumber\\
   &=& (1+Y_\mathrm{i})n_\mathrm{H}k_\mathrm{B}T \simeq \frac{(1+Y_\mathrm{i}) \rho}{m_\mathrm{i}} k_\mathrm{B}T.
\end{eqnarray}
The subscripts `n, i, e, H' refer to neutral hydrogen atoms, ions, electrons and the total number of hydrogen particles (neutral and ionized), respectively, such that the number densities are related by $n_\mathrm{H}=n_\mathrm{n}+n_\mathrm{i}=n_\mathrm{n}+n_\mathrm{e}$.

In computing the radiative loss we follow \citet{1990ApJ...358..328G} and set
\begin{eqnarray}
 \mathcal{L}_\mathrm{rad} &=& -1.547\times 10^{-42}\, Y_\mathrm{i0} \left(\frac{\rho}{m_\mathrm{i}}\right)^2 \alpha T^{1.5},
\end{eqnarray}
where the subscript `0' represents the initial value at time $t=0$. The coefficient $\alpha$ in the above function is the same as the one in Equation~(11) in \citet{1990ApJ...358..328G}.
Their model is applicable up to $T\sim10^5$~K and is well supported by recent investigations of the active solar chromosphere (e.g., Fang, Chen, \& Ding 2003; Jiang, Fang, \& Chen 2010; Xu et al. 2011). We assume the plasma to be in ionization equilibrium. The cooling processes considered in this model \citep{1976ApJ...204..290R} are permitted, forbidden and semiforbidden line transitions, including contributions from dielectric recombination and bremsstrahlung, radiative recombination, and two-photon continua.
The heating function is chosen to initially balance the radiative loss exactly,
\begin{eqnarray}
 \mathcal{H}&=& 1.547\times10^{-42}\, Y_\mathrm{i0} \left(\frac{\rho}{m_\mathrm{i}}\right)^2 \alpha T_0^{1.5}.
\end{eqnarray} 

The simulation domain extends from $x=0$ to $x=L_0$ in $x$ direction and from $y=0$ to $y=2L_0$ in $y$ direction, with $L_0=10^6$~m. Open boundary conditions are used in both $x$ and $y$ directions. 

Three models of the initial current sheet are simulated in this work, differing in the guide field strength and in the inclusion of a density stratification and gravity.  For each model we consider three cases, Cases~A--C, D--F, and G--I in Models~I, II, and II, respectively. In Cases~A, D and G radiative cooling, heating, and ambipolar diffusion are excluded. Cases~B, E and H include heating and radiative cooling but not ambipolar diffusion. Cases~C, F and I include ambipolar diffusion but not the heating and radiative cooling terms. The other initial and boundary conditions and the physical parameters are identical within the cases for each model.

The computations are performed using the MHD simulation code NIRVANA \cite[version 3.6;][]{2011JCoPh.230.1035Z}. This code does not yet have the capability to describe ionization and recombination, so we work here with a non-time dependent ionization ratio $Y_\mathrm{i0}$ and defer the inclusion of these effects to a follow-up investigation.

Adaptive mesh refinement (AMR) is applied. We start the simulation from a base-level grid of $160 \times 320$.  The highest refinement level is 10, which corresponds to a grid resolution $\Delta x \approx 6.1$~m. This resolution is comparable to the neutral-ion collisional mean fee path $\lambda_\mathrm{ni}$ in the middle solar chromosphere (see Table~\ref{Parameter}). Convergence studies have been carried out by repeating some simulations with both a lower and a higher  resolution, with the highest refinement level respectively limited to 9 and 11. The numerical results in the lower resolution case and the higher one are very similar.

\begin{table*}
 \caption{Important parameters of the current sheet region in Models~I,  II and III. Initial values of $T_0$ -- temperature, $S_0$ -- Lundquist number, $n_\mathrm{H0}$ -- hydrogen density, $n_\mathrm{i0}$ -- ion density, $v_\mathrm{A0}$ -- Alfv\'en velocity,  $d_\mathrm{i0}$ -- ion inertial length,  $\lambda_\mathrm{ni0}$ -- neutral-ion collision mean free path, and current sheet width $\delta_\mathrm{ons}$ measured just before secondary islands appear. The height dependence of these parameters is significant only in Model~III.}

  \label{Parameter}
  \begin{tabular}{lcccccccc r@{.}l c}
    \hline
                        &    $T_0$(K) &  $S_0$ & $n_\mathrm{H0}$(m$^{-3}$) &   $n_\mathrm{i0}$(m$^{-3}$) &   $v_\mathrm{A0}$(m/s) &$d_\mathrm{i0}$(m) & $\lambda_\mathrm{ni0}$(m) &    $\delta_\mathrm{ons}$(m)   \\
    \hline
    Model I      &$7000$ &$1.88\times 10^6$&$4.1\times 10^{19}$&$2.3\times 10^{17}$  &$3.32\times10^4$&$0.46$& $4.68$& $400$  \\ 
    Model II     &$7000$ &$5.66\times 10^5$&$4.5\times 10^{20}$&$7.5\times 10^{17}$  &$1.00\times10^4$&$0.24$&  $0.88$ & 80--1200 \\  
    \hline   
     Model III\\
     ($y=L_0$)    &$7000$ &$3.64\times 10^5$&$9.04\times 10^{20}$&$1.05\times 10^{18}$  &$7.28\times10^3$&$0.22$&  $1.19$ & 400 \\                     
     ($y=1.5L_0$) &$7000$ &$1.18\times 10^6$&$8.59\times 10^{19}$&$3.23\times 10^{17}$  &$2.36\times10^4$&$0.39$ &  $3.88$ & 400 \\
     ($y=2L_0$)    &$7000$ &$3.83\times 10^6$&$8.17\times 10^{18}$&$9.06\times 10^{16}$  &$7.66\times10^4$&$0.72$&  $12.58$ & 400 \\
    \hline
  \end{tabular}
\end{table*}

\subsection{Models I and II}\label{ss:models1+2}
Since the effects of radiative cooling and ambipolar diffusion on the reconnection rate depend strongly on the guide field \citep{2011PhPl...18d2105U, 2003ApJ...590..291H}, we consider the Harris sheet with strong guide field (Model~I) and without a guide field (Model~II). These are realized by employing two versions of the Harris current sheet equilibrium, both negelecting gravity and oriented vertically. A force-free Harris current sheet, which is appropriate in a low-$\beta$ environment and has a strong guide field by its nature, is used as Model~I,
\begin{eqnarray}
  B_{x0}&=&0                              \\
  B_{y0}&=&b_0\tanh[(x-L_0/2)/0.05L_0]    \\
  B_{z0}&=&b_0/\cosh[(x-L_0/2)/0.05L_0],
\end{eqnarray}
where $b_0=0.01$~T. The initial current sheet width thus is $\delta_0=0.1L_0$. Due to the force-freeness and neglect of gravity, the initial equilibrium thermal pressure is uniform. The initial plasma $\beta$ is also uniform and set to $\beta_0=0.1$. This yields the initial plasma thermal pressure $p_0 = 12.5\pi^{-1}$~Pa.

To realize a model without guide field, we must deviate from force freeness. Model~II uses a standard Harris current sheet in equilibrium with a pressure gradient,
\begin{eqnarray}
  B_{x0}&=& 0                                         \\
  B_{y0}&=& b_0\tanh[(x-L_0/2)/0.05L_0]            \\
  B_{z0}&=& 0\,.
\end{eqnarray}

From equation~(2), the initial plasma pressure in Model~II is
 \begin{eqnarray}
  p_0 &=& \left(1+\beta_0-\left(\tanh[(x-L_0/2)/0.05L_0]\right)^2\right.)\frac{b_0^2}{2\mu_0},
 \end{eqnarray}
and we set the initial asymptotic plasma $\beta$ also to $\beta_0(|x|\to\infty)=0.1$.
 
In all cases we assume the initial equilibrium temperature as
 \begin{eqnarray}
  T_0 &=& T_{0b}\left[1+(1+\tanh(12(y-L_0)/L_0))/2\right],
  \end{eqnarray}
where $T_{0b}=3500$~K. The functions $p_0(x)$ and $T_0(y)$ yield the total mass density $\rho_0(x,y)$ and particle number density $n_\mathrm{H0}+n_\mathrm{e0}$ for each case. Then we calculate the initial ionization degree $Y_\mathrm{i0}=n_\mathrm{e0}/n_\mathrm{H0}$ using equation~(12) in \citet{1990ApJ...358..328G}. Our settings yield rather similar values of the initial gas pressure, mass density, field strength, and Alfv\'en velocity $v_\mathrm{A0}=b_0(\mu_0\rho_\mathrm{H})^{-1/2}$ in the inflow region ($x=0$ and $L_0$, and $y=L_0$) for both models. However, in Model~II the gas pressure and mass density in the current sheet region are higher by a factor~11 compared to the inflow region. The resulting Alfv\'en velocity in the current sheet is lower than in Model~I by a factor $\approx3$. We should point out that the Alfv\'en velocity calculated in this work is based on the total plasma density, as is appropriate for strongly coupled partially ionized plasmas (at scales $\delta\gg \lambda_\mathrm{ni}$); see, e.g., \citet{1989ApJ...340..550Z}.

The basic plasma parameters in the current sheet region are compiled for the two models in Table~\ref{Parameter} (along with the values of the resulting Lundquist number based on $L_0$, Alfv\'en velocity, and length-scales); they correspond to the conditions in the middle chromosphere, i.e., at roughly 1~Mm above the photosphere, with the field strength being representative of active regions. The neutral-ion collisional mean free path is given by the ratio of the thermal velocity of the neutrals and the neutral-ion collision frequency, 
$\lambda_\mathrm{ni}=v_{T_\mathrm{n}}/\nu_\mathrm{ni}$, with $v_{T_\mathrm{n}}=\sqrt{2k_\mathrm{B}T_\mathrm{n}/m_\mathrm{n}}$, 
$\nu_\mathrm{kl}=n_\mathrm{l}\sigma_\mathrm{kl}\sqrt{8k_\mathrm{B}T_\mathrm{kl}/(\pi m_\mathrm{kl})}$,
$ T_\mathrm{kl}=(T_\mathrm{k}+T_\mathrm{l})/2$, $m_\mathrm{kl}=m_\mathrm{k} m_\mathrm{l}/(m_\mathrm{k}+m_\mathrm{l})$, (where the subscripts k and l denote the species),  and the value of the collisional cross section $\sigma_\mathrm{kl}$ is taken from \citet{2012ApJ...747...87K}.

Initial perturbations of both magnetic field ($\bm{B}^{\prime}$, with $\nabla \cdot \textbf {B}^{\prime}=0$) and 
velocity ($\bm{v}^{\prime}$) are applied at $t=0$ in all cases to trigger the reconnection process. 
The perturbations result in a thinning of the current sheet in two sections between a set of three primary islands, whose midpoints are located at $y=0$, $L_0$, and $2L_0$ (see Fig.~1(a)). In this paper, we will focus only on the section in the domain $L_0<y<2L_0$, i.e., the bottom half of the box is used as an auxiliary part of the computation only. Its function is to yield a stationary primary plasmoid at the bottom of the height range of interest, which is not influenced by any effects of a numerical boundary; the primary plasmoid acts like a line-tied bottom of the current sheet of interest. The initial temperature $T_0$ varies from 5250~K to 7000~K in the domain of interest, but it quickly increases to the upper value above the stationary plasmoid at $y=L_0$, such that essentially all of the upper current sheet initially is at $T_0\approx7000$~K. The resulting ionization degree in the current sheet is $0.5\%$ in Model~I and $0.2\%$ in Model~II. 

We use a simple parametrization of the magnetic diffusion which relatively closely matches the diffusion computed from a model atmosphere in \citet{2012ApJ...747...87K}, $\eta=5\times 10^4 (3500/T)^{1.5}$~m$^2$\,s$^{-1}$. This yields the initial diffusion in the current sheet as $1.77\times10^4$~m$^2$\,s$^{-1}$. The Lundquist number based on this diffusivity, $v_\mathrm{A0}$, and $L_0$, which corresponds to the `global scale' of the current sheet in the upper part of the box, is $S_0= 1.88 \times 10^6$ in Model~I and $ 5.66 \times 10^5$ in Model~II. $\eta$ decreases as the temperature increases. From the numerical results presented in the following section, the value of $\eta$ at the main $X$ point drops to $\sim2000$~m$^2$\,s$^{-1}$ during the secondary instability process, and the Lundquist number then reaches $\sim10^7$.

The ambipolar diffusion coefficient $\eta_\mathrm{AD}$ is defined as \citep{1965RvPP....1..205B}
 \begin{eqnarray}
  \eta_\mathrm{AD}&=& \left(\frac{\rho_\mathrm{n}}{\rho}\right)^2(\rho_\mathrm{i}\nu_\mathrm{in}+\rho_\mathrm{e}\nu_\mathrm{en})^{-1},
 \end{eqnarray}
where, respectively, $\nu_\mathrm{in}$ and $\nu_\mathrm{en}$ are the ion-neutral and electron-neutral collision frequencies defined analogous to $\nu_\mathrm{ni}$ above,
with the cross sections $\sigma_\mathrm{in}=5 \times 10^{-19}$~m$^2$ and $\sigma_\mathrm{en} = 10^{-19}$~m$^2$ again taken from \citet{2012ApJ...747...87K}. This yields
 \begin{eqnarray}
    \eta_\mathrm{AD}&=& 1.65 \times 10^{-11}
  (\frac{1}{Y_i}-1)\frac{1}{\rho^2\sqrt{T}}
                        ~~\mbox{m$^3$\,skg$^{-1}$},
 \end{eqnarray}
 with the temperature inserted in Kelvin. The ambipolar diffusion coefficient is a function of ionization degree, temperature and plasma density. According to the model atmospheres by \cite{2008SoPh..251..589K} and \cite{1986A&A...154..231P}, the magnitude of $\eta_\mathrm{AD}$ varies from $10$ to $10^5$~m$^3$\,s\,kg$^{-1}$ from the bottom to the top of the chromosphere. Here we set $\eta_\mathrm{AD}=5\times10^4(T_0/T)^{1/2}(\rho_0/\rho)^2$~m$^3$\,s\,kg$^{-1}$. This is representative of the ambipolar diffusion coefficient in the middle chromosphere.

For the radiative loss and heating function in Models~I and II, we simplify the coefficient $\alpha$ compared to the expression in \citet{1990ApJ...358..328G}, which is slowly varying in the middle chromosphere (about 1000~km above the photosphere), the height range of interest in the present study. Since we do not investigate the height dependence of the reconnection rate, we here use a representative uniform value of $\alpha=0.01$, which is chosen such that the resulting magnitude of the initial radiative loss is of order $0.07\mbox{--} 3.5$~J\,m$^{-3}$\,s$^{-1}$, consistent with their results.

\subsection{Model III}\label{ss:model3}

Since the active section of the current sheet in our computations, $L_0\lesssim y<2L_0$, is of order 1~Mm, it encompasses a considerable range of density and, correspondingly, Alfv\'en velocity in the chromosphere. In Model~III we therefore include gravity and the resulting stratification of thermal pressure and plasma density. It is not possible to construct an equilibrium consisting of a vertical Harris sheet without a guide field in the presence of gravity. Therefore, Model~III combines the force-free Harris sheet of Model~I with density stratification, gravity, and a balancing pressure gradient, given by
  \begin{eqnarray}
    \partial_x p_0(y) &=& 0,  \\
    \partial_y p_0(y) &=& -273.9\rho_0(y), \\
    p_0(y) &=& \frac{(1+Y_\mathrm{i0}) \rho_\mathrm{H0}(y)}{m_\mathrm{i}} k_\mathrm{B}T_0,
 \end{eqnarray}
where, for simplicity, the initial temperature $T_0=7000$~K is chosen to be uniform, and $\rho_\mathrm{H0}=n_\mathrm{H0}m_i$. According to Equation~(12) in \citet{1990ApJ...358..328G}, the ionization degree is simplified as
 \begin{eqnarray}
   Y_\mathrm{i0}=\frac{n_\mathrm{e0}}{n_\mathrm{H0}} \approx \sqrt{\frac{\phi_0}{n_\mathrm{H0}}},
 \end{eqnarray}
where $\phi_0 \simeq 1.21 \times 10^{15}$~m$^{-3}$ and $\phi_0 \ll n_\mathrm{H0}$ in our model. From Equations~(22--25), we obtain
 \begin{eqnarray}
  n_\mathrm{H0} \simeq n_\mathrm{H00}\exp\left(-\frac{4.706y}{L_0}\right), 
 \end{eqnarray}
where we assume $n_\mathrm{H00}=10^{23}$~m$^{-3}$. The same perturbations as those in Models~I and II are applied; we also only focus on the section in the domain $L_0 < y< 2L_0$.  From Equations~(24--26), we calculate the initial distributions of hydrogen number density $n_\mathrm{H0}$, ion number density $n_\mathrm{i0}$, ionization degree $Y_\mathrm{i0}$, Alfv\'en velocity $v_\mathrm{A0}$, ion inertial length $d_\mathrm{i0}$, and neutral-ion collision mean free path $\lambda_\mathrm{ni0}$. The values of these parameters in Model~III at $y=L_0$, $y=1.5L_0$ and $y=2L_0$ are compiled in Table~\ref{Parameter}.

The initial Lundquist number $S_0=Lv_\mathrm{A0}/\eta$ is of order $10^6$ (see Table~\ref{Parameter}). The initial ambipolar diffusion coefficient takes values of $\eta_{AD}=75.67$~m$^3$\,s\,kg$^{-1}$ at $y=L_0$, $2.57\times10^3$~m$^3$\,s\,kg$^{-1}$ at $y=1.5L_0$, and $8.71\times 10^4$~m$^3$\,s\,kg$^{-1}$ at $y=2.0L_0$.
 
Since the density stratification in $y$ direction is included in Model~III, we here use the full $y$-dependent expression for $\alpha$ from \citet{1990ApJ...358..328G}, so that the resulting height dependence of the radiative loss is similar to their results. Compared to their expression, we here increase the value of $\alpha$ by a factor of 10, such that the initial radiative loss is in the range $\approx(0.2\mbox{--}1.3)$~J\,m$^{-3}$\,s$^{-1}$ for $L_0<y<2L_0$, considerably higher than the value of $\approx0.07$~J\,m$^{-3}$\,s$^{-1}$ in Model~I. This choice is made to demonstrate that a main result we will obtain for Model~I, namely that radiative losses in the chromosphere do not significantly influence the evolution of current sheets with a strong guide field, remains valid for the much higher level of radiative losses. The levels chosen for the initial radiative losses in Models~I and III both fall in the range deduced in \citet{1990ApJ...358..328G}.

From the above descriptions, we find that model~III is close to the conditions in the middle chromosphere, i.e., in the height range $\approx(0.5\mbox{--}1.4)$~Mm above the photosphere. Figure~1(b) displays the initial magnetic field and plasma density.

\section{Results}\label{s:results}
\subsection{Cascading plasmoid instability with internal slow-mode shocks}\label{ss:cascading}

The Lundquist number in our models exceeds the critical Lundquist number for onset of the plasmoid instability substantially. As the Harris sheet evolves to a thinner current sheet in response to the initial perturbation, it first passes through a Sweet-Parker-like phase (with the exception of Case~F) as indicated by its smooth, very elongated structure (which contains an $X$-point at $y\approx1.5~L_0$) and an average reconnection rate $\gamma\sim10^{-3}$ in agreement with the Sweet-Parker scaling (see below for the method of quantitative analysis). When the current sheet sufficiently thins such that a critical aspect ratio is reached, the plasmoid instability sets in, forming multiple islands. Subsequently, the instability cascades to smaller scales. Throughout this process a high reconnection rate, rapidly fluctuating in the range $\sim0.01\mbox{--}0.03$, is realized (see Sect.~\ref{ss:analysis}). The plasmoid cascading process is found to be very similar in all nine cases run.

Fig.~2 displays the plasmoid cascading process for Model~III, Case~I with ambipolar diffusion included. The left panel shows the primary island at $y=L_0$ produced by the initial perturbation and a number of secondary islands formed by the plasmoid instability (note that the primary island at $y=2L_0$ was ejected by the slow reconnection outflows that evolved in the Sweet-Parker-like phase). As we zoom in (panels~2-3), several third-order plasmoids become visible in the current sheet, which has further narrowed between the secondary plasmoids. Zooming in further (panels~3-4), fourth-order plasmoids can be seen clearly. This plasmoid cascading process is very similar to the results of \cite{2011ApJ...737...24B}, who have simulated it for coronal parameters. It is found here for the first time in partially ionized plasma. 

The half-width of the thinnest fragment current sheet in Fig.~2 is about 25~m, clearly larger than the neutral-ion collisional mean free path $\lambda_\mathrm{ni}$ (Table~\ref{Parameter}), thus consistent with the one-fluid approximation. The cascading plasmoid instability occurs in all nine runs considered in this paper, reaching comparable reconnection rates in the cases with and without a guide field, which differs strongly from the results for stationary reconnection in partially ionized plasma \citep{2003ApJ...590..291H, 2011PhPl...18d2105U}. The gravity of the Sun also has only little effect on the time dependent reconnection rate (see below).

Given the large body of experience from one-fluid, Hall-MHD and kinetic models of reconnection in the plasmoid-unstable range \cite[e.g.,][]{2011PhPl...18g2109H, 2013arXiv1306.1050M, 2009PhRvL.103f5004D}, one can reasonably expect that the plasmoid cascading process tends to continue to even smaller scales than resolved here, down to the ion inertial length $d_\mathrm{i}$. In partially ionized plasma the process is modified by recombination within the plasmoids, reducing their current and flux \citep{2012ApJ...760..109L}, and it is not yet known how this influences the cascading process. A less dynamical evolution of the plasmoids is conceivable, but since the cascading occurs between previously formed plasmoids, it appears unlikely that recombination within plasmoids would suppress it completely. This is indeed indicated by the results in \citet{2013PhPl...20f1202L}, who find that that the plasmoid instability can raise the reconnection rate also beyond the validity range of the one-fluid approximation. We thus expect that the multiple levels of the plasmoid instability process connect the global, large MHD-scale reconnection with the local, small kinetic-scale reconnection also in partially ionized plasma. It is well known that the plasmoid instability in fully ionized plasma yields a high reconnection rate throughout the cascading process, and our simulations demonstrate that the cascading plasmoid instability operates in weakly ionized plasma in the one-fluid limit in basically the same manner as in fully ionized plasma, reaching a high reconnection rate as well.

As is well known, reconnection at $X$-points always involves slow-mode shocks in the Petschek regime. In our simulations, we find that many small-scale slow-mode shocks form transiently at the edge of secondary plasmoids, attached to the neighboring $X$-point in a secondary fragment of the current sheet. Fig.~3 shows a pair of such shocks. Sudden changes of the magnetic field direction and current density at the left and right edges of the plasmoid can clearly be seen; these are very similar to the slow-mode shock structure displayed, e.g., in \cite{1988SoPh..117...97F}, \cite{1997AdSpR..19.1797S} and \cite{2012MNRAS.425.2824M}. Fig.~3(b) shows that the temperature increases strongly not only within the current sheet fragment but also downstream of the slow-mode shocks in the plasmoid. The angle between the shock front and the $y$-direction is $\approx5.4^{\circ}$. Fig.~4 displays the magnetic field and the current density along a cut in the $x$-direction at $y=1.691~L_0$. One can see that the magnitude of the field component tangential to the shock, $B_{\parallel}$, decreases rapidly toward the downstream side and that the current density $J_z$ has a peak where $B_{\parallel}$ changes rapidly, but the component normal to the shock, $B_{\perp}$,  stays nearly uniform along the cut. This is exactly the behavior of a slow-mode shock. Similar structures exist in the outflow region of the current sheet. These slow-mode shocks could be a candidate mechanism to explain the chromospheric heating \citep{2011PlPhR..37..161K}.

\subsection{Effects of radiative cooling and ambipolar diffusion}\label{ss:analysis}
\subsubsection{Strong guide field}

We now compare the simulation results of the three cases which were run for Model~I. The reconnection rate is computed as the rate of change of the magnetic flux accumulated between the $O$-point in the primary island at $y=L_0$ and the main reconnection $X$-point (see Ni et al. 2012; Ni et al. 2013),
 \begin{equation}
     \gamma(t)=\frac{\partial(\psi_X(t)-\psi_O(t))}{\partial t}\frac{1}{b_0 v_{A0}}.
     \label{e:rec_rate}
 \end{equation}              
Here the flux function $\psi$ is defined through the relations $B_x = - \partial \psi/ \partial y$, $B_y = \partial \psi/\partial x$, and the main reconnection $X$-point is determined as the $X$-point which has the highest $\psi$ value of all $X$-points in the box. This analysis is performed on the relatively low grid refinement level~3 which ensures uniform coverage of the current sheet section from the primary $O$-point to the upper boundary throughout the runs. This choice still captures much of the temporal variability of the reconnection rate. Local quantities like the current density, temperature, and current sheet width at the main reconnection $X$-point are determined at the highest refinement level chosen by the code, to evaluate their full dynamic range; after the onset of the plasmoid instability this is level~10 in all nine cases. The current sheet width is determined as the full width at half maximum (FWHM) of the current density profile in $x$ direction.

\begin{figure*}
\centerline{\includegraphics[width=0.40\textwidth, clip=]{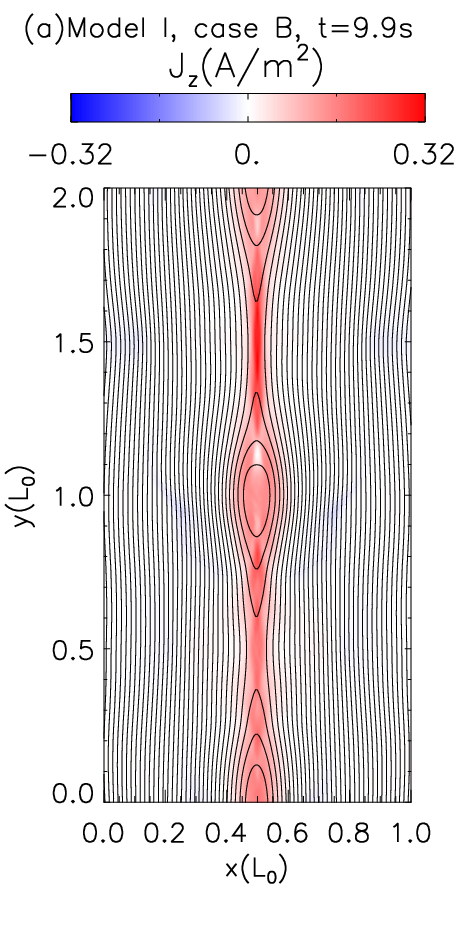}
                      \includegraphics[width=0.40\textwidth, clip=]{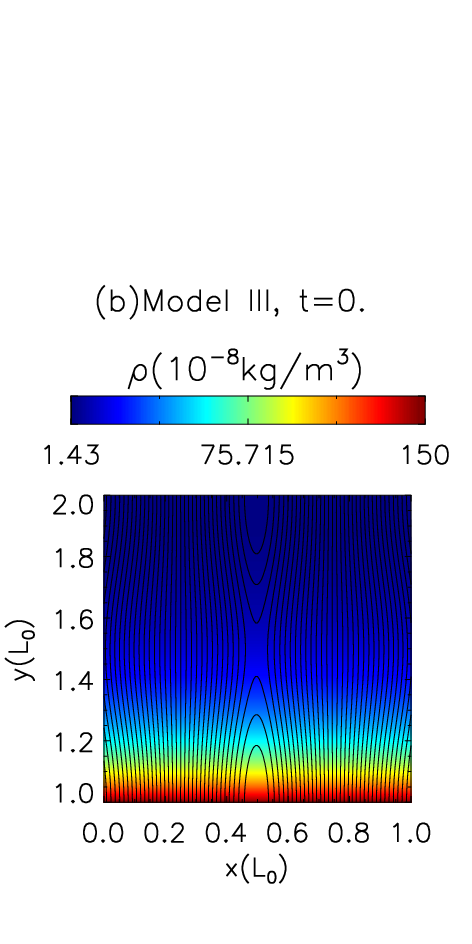}}
    \caption{(a) Field lines and current density $J_z$ (background color) in the whole simulation domain for the strong guide field model~I with radiative cooling (Case~B) at $t=9.9$~s.
             (b) Field lines and plasma density $\rho$ (background color) in the domain we focus on for Model~III with gravity at $t=0$ (with the initial perturbation included).}
\label{fig.1}
\end{figure*}

The reconnection rate for Cases~A--C is displayed in Fig.~5(a). The three cases are very similar to each other and show a dynamical behaviour that is basically analogous to reconnection in fully ionized plasma, as expected from our one-fluid approximation, especially for Case~A in the absence of heating, radiative cooling, and ambipolar diffusion. Up to $t\approx26$~s the reconnection rate remains at a low level, reaching the Sweet-Parker value of $\gamma \sim S^{-1/2}\sim10^{-3}$ for $t\ga20$~s. At this time the decreasing current sheet width has also reached the Sweet-Parker value $\delta x \sim S^{-1/2}L_0 \sim 10^3$~m. This appears to be a Sweet-Parker reconnection phase, which is also supported by the elongated, smooth appearance of the current sheet (see Fig.~\ref{fig.1}).

Subsequently, a fast rise of the reconnection rate commences, which quickly saturates at a level $\gamma\approx0.02$. The first secondary islands due to the plasmoid instability have clearly developed by the end of the fast rise, which we thus interpret as the linear phase of the plasmoid instability. All quantities show the high variability characteristic of the plasmoid instability in the further evolution. The current sheet width, measured at the main $X$-point, gradually decreases in the Sweet-Parker phase and enters a rapid decrease down to $\delta x \approx 30$~m simultaneously with the onset of the plasmoid instability (Fig.~6(a)). This minimum current sheet width is comparable to the grid scale at refinement level~10 ($\delta x \approx 5\Delta x$) but still much larger that the neutral-ion collisional mean free path and the ion inertial length (Table~1), so that our one-fluid approximation is valid throughout the computation. The current density at the main $X$-point shows a rather close inverse proportionality to the current sheet width. The average reconnection rate remains at a high level throughout the cascading phase of the instability. It shows a trend of gradual decrease to an average level of $\approx0.015$ which is most likely due to the shortening of the active section of the current sheet as the primary island at $y=L_0$ grows; the field lines in the inflow region must then bend increasingly before reconnecting at the $X$-point. The temperature at the main $X$-point increases rapidly from the onset of the plasmoid instability, saturating at $\sim3\times10^4$~K (Fig.~7(a)). The corresponding decrease of the magnetic diffusivity raises the effective Lundquist number in the course of the plasmoid instability by nearly an order of magnitude. It should be noted that the peak temperature in the current sheet, located within the islands downstream of the slow-mode shocks, continues to rise, reaching $\sim8\times10^4$~K during the plasmoid instability.

The comparison of Cases~A and B shows that the radiative cooling hardly has any effect on the evolution of the reconnecting current sheet at the chosen parameters of the middle chromosphere if the sheet includes a considerable guide field. This is fully in agreement with earlier investigations \cite[e.g.,][]{2011PhPl...18d2105U}, which had pointed out that a sufficiently strong guide field suppresses the current sheet thinning due to radiative cooling by making the sheet incompressible. The only remaining effect on the reconnection rate is given by the temperature dependence of the magnetic diffusivity. \citet{2011PhPl...18d2105U} considered the case that radiative cooling in a stationary Sweet-Parker sheet with negligible heating increases the reconnection rate in this way. However, heating is a key factor of the energy balance in the chromosphere, and so our model also includes a heating term which is adjusted to match the initial cooling rate and to have the same dependence on the density. Consequently, the radiative losses will not cool the plasma below the initial temperature $T_0$; they have a strong effect only when the temperature increases considerably above $T_0$. Lower temperatures can be reached by adiabatic cooling if the plasma expands (an effect seen below in Fig.~\ref{fig.9}), but this is not the case here at the $X$-point, where the temperature has the same value as in Case~A during the Sweet-Parker phase (Fig.~7(a)). The subsequent fast reconnection dominated by the plasmoid instability is largely independent of the magnetic diffusivity \citep{2010PhPl...17f2104H}. Temperatures of $(4\mbox{--}5)T_0$ are reached in this phase, as in Case~A, which shows that the strong heating at the slow-mode shocks dominates the radiative losses in this phase.

Considering the effect of ambipolar diffusion in Case~C, again only a weak influence on the reconnection rate is found. The only differences to Cases~A and B are a more impulsive and slightly earlier onset of the plasmoid instability and a somewhat higher variability near the end of the Sweet-Parker phase. We interpret this behaviour as follows. In the Sweet-Parker phase a strong thinning of the current sheet is inhibited by the guide field \citep{2003ApJ...590..291H}. The ambipolar electric field nevertheless increases strongly as the sheet thins in response to the initial perturbation. However, the area of enhanced $E_\mathrm{AD}$ does not reach the plane $x=0.5~L_0$ where the antiparallel field component reverses sign, but rather forms two layers on either side of of the field reversal plane (Fig.~\ref{fig.11}(a), (c)). Thus, it does not amplify the annihilation of the field in this plane. The width of the layers, $\sim500$~m, agrees with the ion-neutral decoupling length scale $L_\mathrm{AD}=v_\mathrm{Ai}/\nu_\mathrm{in}$, where $v_\mathrm{Ai}$ is the Alfv$\acute{e}$n velocity based on the ion density \citep{2011PhPl...18k1211Z}. As a dominant $X$-point develops, the associated reconnection flows convect the two layers with enhanced ambipolar electric field to the field reversal plane (Fig.~\ref{fig.11}(b), (c)), so that it now contributes to the change of flux (Equation~\ref{e:induction}). The comparison of Figs.~\ref{fig.11}(a) and (b) shows that the rate of change of the magnetic field in the Sweet-Parker-like current sheet is dominated by the resistive contribution to the electric field, whereas it is dominated by the ambipolar electric field during the rapid rise of the reconnection rate at the first $X$-point of the commencing plasmoid instability. The layers of enhanced ambipolar field intermittently approach the field reversal plane already during the considerable thinning of the current sheet in the final part of the Sweet-Parker phase (Fig.~6(a)), causing the intermittent moderate enhancements of the reconnection rate seen in Fig.~5(a). Since the inclusion of the ambipolar diffusion terms causes a strong decrease of the adaptive time step, this run was terminated after the reconnection rate reached the saturation level.

The three cases also show a very similar current sheet aspect ratio at the onset of the plasmoid instability. We refer to this quantity as the ``onset aspect ratio'' and determine it from the current sheet width $\delta x$ and the similarly defined FWHM length of the current sheet in $y$ direction, $\delta y$. We find $\delta x\approx400$~m and $\delta y\approx3.5\times10^5$~m, i.e. an onset aspect ratio $\delta_\mathrm{ons}=\delta y/\delta x\approx 875$ in all three cases (see Fig.~8 for Case~B). This value lies considerably above the critical aspect ratio for onset of the plasmoid instability at the point of transition between the Sweet-Parker and plasmoid-mediated reconnection regimes, $S_\mathrm{cr}^{1/2}$, when the Lundquist number is varied to reach the critical value $S_\mathrm{cr}\sim10^3\mbox{--}10^4$ \citep{2010PhPl...17f2104H, 2012PhPl...19g2902N, 2012ApJ...760..109L}. Our simulations are performed with far higher Lundquist numbers in the range $S\sim10^6\mbox{--}10^7$ (varying in time due to the evolving $\delta x$ and $T$), and correspondingly the current sheet aspect ratio rises considerably above the critical value $S_\mathrm{cr}^{1/2}$, roughly aproaching $S^{1/2}$.

In summary, in current sheets with a strong guide field, the plasmoid instability mediates a fast reconnection regime in partially ionized plasma, while radiative cooling and ambipolar diffusion have only very little effect. For the parameters of the middle chromosphere, the reconnection rate is enhanced by a factor $\sim20$ above the Sweet-Parker rate, reaching the range of observationally inferred values.

\subsubsection{Zero guide field}

Next we consider the Harris current sheet with vanishing guide field (Model~II, Cases~D--F). A major difference to all corresponding runs for Model~I is the much slower evolution of the current sheet, due to our choice to realize the necessary pressure enhancement in the sheet by a density enhancement and the resulting smaller Alfv\'en velocity in the sheet. The reconnection rate, which is normalized to the field strength and Alfv\'en velocity in the inflow region, thus takes a much longer time to rise. The density in the current sheet gradually decreases during our runs, and eventually the Alfv\'en velocity in the sheet becomes comparable to the external Alfv\'en velocity. The reconnection rate then reaches similar values as in Model~I. 

In all three cases run for Model~II, reconnection rates exceeding the Sweet-Parker value develop only from the onset of the plasmoid instability (Fig.~\ref{fig.8}), which happens at $t\approx92$, 84, and 50~s for Cases~D, E, and F, respectively (Figs.~5(b) and 8). A Sweet-Parker-like regime develops only transiently, for $\sim10$~s prior to the onset of the plasmoid instability, in Cases~D and E as the current sheet width approaches the Sweet-Parker value (Fig.~6(b)). As for Model~I, the onset of the plasmoid instability is associated with a rapid decrease of the current sheet width. Due to the high inertia of the plasmoids in the initially dense current sheet, the instability shows a far more gradual overall development compared to the strong guide field cases.

The reconnection rate in Case~D without radiative cooling and ambipolar diffusion does not reach saturation but rather rises essentially monotonically (apart form the small time-scale fluctuations characteristic of the instability) until the main $X$-point leaves the box through the upper boundary at $t\approx230$~s. A peak reconnection rate of $\gamma\approx0.025$ is then reached; it would be somewhat higher (lower) if the box were chosen somewhat larger (smaller). The current sheet width continues to decrease in the course of the cascading plasmoid instability until it reaches $\delta x\approx30$~m, as in Model~I. The temperature at the $X$-point reached at the end of the run, $T\sim3\times10^4$~K, is also similar to the result for Model~I.

The evolution in Case~E with radiative cooling and heating is qualitatively similar to Case~D, but it commences earlier and is more impulsive. Here the current sheet width drops to $\delta x\approx30$~m, and the reconnection rate rises to $\gamma\approx0.025$, in about one half of the time. This more dynamic behaviour is related to a faster decrease of the density in the current sheet after $t\approx100$~s. It appears that the reconnection rate saturates at this level until it sharply drops as the main $X$-point leaves the box at $t\approx170$~s. The effect of the radiative cooling again remains minor prior to the onset of the plasmoid instability. The current sheet is then only slightly thinner than in Case~D (Figs.~6(b) and 8). The effect of radiative cooling becomes more obvious at the high temperatures in the course of the plasmoid instability, limiting the temperature to about half the peak value found in Case~D (Fig.~7(b)). Radiative cooling in Case~D acts more efficiently than in Case~B because here the densities are higher.

As expected in the absence of a guide field, the inclusion of ambipolar diffusion in Case~F allows the current sheet to thin very efficiently \citep{1994ApJ...427L..91B}. The evolution here begins with a rapid decrease of the current sheet width at the main X-point down to $\delta x\approx30$~m (within $\approx10$~s), which is near its smallest value observed in all runs at the AMR refinement levels 10 and 11. A thinning by a factor 500 is expected from the ratio of ion and neutral pressure (the inverse ionization degree for our isothermal conditions) when only the ion pressure in the sheet balances the magnetic pressure of the external region. The actual thinning goes beyond that by a factor $\sim3$, but this already includes the onset of the plasmoid instability at $t\approx50$~s and $\delta x\approx80$~m. The instability seamlessly grows out of the current sheet collapse mediated by the ambipolar diffusion, without leaving room for a Sweet-Parker-like phase, since the onset aspect ratio is obviously reached during the collapse. The reconnection rate grows in a manner similar to Cases~D and E, at least to a level $\gamma\sim0.02$. It is possible that it would grow to higher values if the run could be continued. However, the main reconnection X-point has moved out of our simulation domain after $t=150$~s. It is seen that the reconnection rate then suddenly decreases to a low value, similar to Cases~E and D.  Overall, the dynamical evolution of the plasmoid instability appears to be similar to Cases~D and E, including the temperatures reached at the main $X$-point (Fig.~7(b)), so that a saturation of the reconnection rate at similar levels is conceivable. 

The current sheet widths and aspect ratios for onset of the plasmoid instability are, approximately, 1200~m and 500 in Case~D, 950~m and 730 in Case~E, and 80~m and 1250 in Case~F (see Fig.~\ref{fig.8}). The onset occurs at a higher aspect ratio when the sheet thins more dynamically.

To summarize the runs for Model~II without guide field, we find a basic analogy to the strong guide field cases in that the plasmoid instability is the dominant mechanism to realize fast reconnection, with rates similar to the strong guide field cases. Although our simulations confirm the strong thinning of the current sheet allowed by ambipolar diffusion and anticipated to yield an accelerated Sweet-Parker-like reconnection regime \citep{2003ApJ...583..229H}, this regime does not occur because the aspect ratio for onset of the plasmoid instability is reached by the end of the ambipolar diffusion-driven current sheet thinning. An accelerated Sweet-Parker-like reconnection due to radiative cooling of the current sheet does not occur in our model because we include a background heating term representing the chromospheric heating.

\subsection{Effects of the density stratification}\label{ss:stratification}

We find that the inclusion of gravity and density stratification in Model~III has surprisingly little effect on the main results found for Models~I and II, although the density varies by two orders of magnitude in the relevant height range $y>L_0$. Specifically, (1) the plasmoid cascading process shown in Fig.~2 is very similar to Models~I and II; (2) the reconnection rate, current sheet width and temperature at the main X-point for Cases~G--I in Model~III behave very similar to those for Cases~A--C in Model~I, respectively (Figures~5--7); (3) many fine slow-mode shock structures attached around the edges of the plasmoids appear, where the highest temperatures are produced; (4) the onset aspect ratios for Models~I and III are the same, as shown in Fig.~8 and Table~1; (5) radiative cooling and ambipolar diffusion have little effect on the plasmoid instability in Model~III; (6) the maximum out-flow velocity in Model~III is the same as in Model~I, about $40$~km\,s$^{-1}$. The similarity of the reconnection rate, which is normalized by the same value of $b_0v_{A0}$ as in Model~I, and other quantities at the main X-point can be understood from the rather similar values of the Alfv\'en velocity at that point in Models~I and III, with the Alfv\'en velocity initially being higher in Model~III by only a factor $\approx1.2$; this implies a similar dynamic behavior in the vicinity of the main X-point.

There are some details in the reconnection process which are different in Models~I and III. As shown in Figs.~8 and 9, the current sheet gradually becomes inclined to the side, especially after secondary islands appear. This is due to gravity acting on the dense plasmoids, which form slightly asymmetrically about the current sheet as a consequence of roundoff errors. Secondary islands in Model~III start to appear later than in Model~I for the same initial perturbation. In the period before secondary islands appear, the hottest plasma exists in the up-flow region in Model~III (Figure~10), but in the down-flow region in Model~I (not shown). Since low-density plasma can be heated more easily, the rapid decrease of the plasma density with increasing height in Model~III is the main reason for this difference. For Case~H, the hottest plasma usually appears around the edge of the plasmoids, but their center is still relatively much cooler (Figure~10), the reason being that the plasma density inside the plasmoids is higher and, therefore, radiative cooling is stronger. During the reconnection process, the main X-point in Model~III gradually moves to a higher place compared to Model~I, as seen in Figures~9 and 12. With radiative cooling, this results in a higher temperature for Model~III ($\sim80000$~K in Case~H vs. $\sim70000$~K in Case~B).

\subsection{Comparison with observations}\label{ss:comparison}

Micro-jets have been frequently observed in the low atmosphere (e.g., Liu et al. 2009; Morton 2012; Singh et al. 2012; Bharti, Hirzberger, \& Solanki 2013). Magnetic reconnection is being considered as one of the mechanisms that may cause the ejection of these jets. Their velocity is usually found in the range of 10--150~km\,s$^{-1}$. The average speed of chromospheric jets around the edges of sunspots is around 30~km\,s$^{-1}$ (Morton 2012). Fig.~\ref{fig.12} shows that the upward outflow velocity $v_y$ can reach about 40~km\,s$^{-1}$ in the strong guide field model with radiative cooling and about 25~km\,s$^{-1}$ in the zero guide field model. These upward reconnection outflow velocities are close to the typical velocities of chromospheric jets. Higher outflow velocities (by a factor $\sim\beta_0^{-1/2}\sim3$) from a chromospheric current sheet with zero or weak guide field can be expected if the sheet is not overdense as in our Model~II. The outflows become intermittent in the course of the plasmoid instability as relatively large merged islands are convected out of the sheet (Fig.~\ref{fig.12}). This corresponds to observations of plasmoids in chromospheric jets \cite[e.g.,][]{2012ApJ...759...33S}.

Many observations demonstrate that chromospheric jets can be heated to the transition region temperature (e.g., Teriaca, Curdt, \& Solanki 2008). \citet{2012A&A...543A...6M} pointed out that ``slower'' ($v \sim 30$~km\,s$^{-1}$) jet events near sunspots involve $10^5$~K plasma and could play a role in heating the chromosphere and corona. Although radiative cooling is included in Cases~B and E, these simulations yield temperatures in the current sheet region of several $10^4$~K (Figs.~3, 7, and \ref{fig.10}), and the plasma immediately downstream of the slow-mode shocks and in turbulent regions between secondary islands is heated to temperatures of the lower transition region, $\sim8\times10^4$~K. Therefore, our simulations demonstrate that magnetic reconnection driven by the plasmoid instability in the low atmosphere contributes to chromospheric heating.

\section{Conclusions and discussion}\label{s:discussion}

Magnetic reconnection in the middle chromosphere, including radiative cooling, heating, and ambipolar diffusion effects, has been studied in the 2.5-dimensional MHD approximation. A Harris current sheet model with either a strong guide field (Models~I and ~III) or vanishing guide field (Model~II) is considered. We have assumed that the plasma consisting of ions, electrons and neutral hydrogen atoms is strongly coupled and in ionization equilibrium state, that the pressure is isotropic, and that the heat flux can be neglected. The ionization and radiative loss model by \cite{1990ApJ...358..328G} is used, resulting in an initial ionization degree in the current sheet in the range $\approx(0.2\mbox{--}0.5)$\,\%. A temperature-dependent magnetic diffusion coefficient adapted to the values in \citet{2012ApJ...747...87K} is employed, leading to Lundquist numbers of $S\sim10^6\mbox{--}10^7$.

The main numerical results can be summarized as follows:

1. The plasmoid instability mediates a fast reconnection regime under chromospheric conditions for vanishing as well as for strong guide field. Reconnection rates of order 0.01--0.03, a factor $\sim20$ above the Sweet-Parker rate and within the range of observationally inferred rates, are reached in either case, based on the classical Spitzer resistivity. The aspect ratio at the onset of the plasmoid instability at the supercritical values of the Lundquist number considered here is found to lie in the range 500--1250.

2. AMR allowed us to resolve three levels of secondary islands and current sheet widths down to $\delta x\approx30$~m, where the one-fluid approximation is still valid.

3. Ambipolar diffusion and radiative cooling have a significant influence on the reconnection process only in the model with vanishing guide field. Ambipolar diffusion then leads to the expected strong current sheet thinning \citep{1994ApJ...427L..91B}, but the current sheet thins very fast, reaching the onset aspect ratio, such that an accelerated Sweet-Parker reconnection regime does not develop, but rather the plasmoid instability sets in. The expected strong current sheet thinning in a Sweet-Parker regime due to radiative cooling \citep{2011PhPl...18d2105U}  does not occur in our simulations because we include a background heating term. This thinning can generally not be expected to be strong in the middle and lower chromosphere, where the temperature is hardly higher than twice the solar temperature minimum. Both ambipolar diffusion and radiative losses have an effect on the plasmoid instability for vanishing guide field, implying, respectively, a much earlier onset and a faster development, but apparently no significant increase in the reconnection rate.

4. Many slow mode shocks are generated between secondary plasmoids and secondary fragments of the current sheet during the unstable reconnection process. Downstream the shock front regions (in the plasmoids), the temperature is highest and reaches $T\sim 8\times10^4$~K, i.e., lower transition region temperatures. This supports conjectures in the literature that fast magnetic reconnection is a candidate mechanism for the chromospheric heating.

5. Upward outflow velocities in the course of the plasmoid instability reach $\approx40$~km\,s$^{-1}$ and $\approx25$~km\,s$^{-1}$ in the strong and zero guide field cases, respectively. Outflow velocities higher by a factor $\sim3$ are expected in the zero guide field case if the initial current sheet were not chosen to be over dense. These values lie in the range observed for chromospheric jets. Dynamic plasmoids yield variations in the outflow, consistent with observations of plasmoids in chromospheric jets.

Recent work by \citet{2012ApJ...760..109L, 2013PhPl...20f1202L} has studied the effects of recombination on reconnection in the middle chromosphere in a two-fluid model (weak coupling between ions and neutrals). This also included the occurrence of the plasmoid instability. Reconnection rates of order 0.1 were found to result from recombination effects, while the plasmoid instability gave only a minor additional increase of $\sim15$\,\%. Recombination within the plasmoids reduces their current, which may slow down the dynamics of the plasmoid instability. It thus appears that recombination must ultimately be included for a complete description of reconnection in partially ionized plasma. However, the small increase of the reconnection rate due to the plasmoid instability in \citet{2012ApJ...760..109L} results in part from their choice of a high magnetic diffusivity, $\eta = 6\times10^4\mbox{--}2.4\times10^6$~m$^2$\,s$^{-1}$, which is about two orders of magnitude higher than a typical magnetic diffusivity in the middle chromosphere and did not change with temperature. Our diffusivity is $\eta_0=1.8\times10^4$~m$^2$\,s$^{-1}$ initially and decreases to $\sim2 \times10^3$~m$^2$\,s$^{-1}$ as the current sheet is heated in the course of the plasmoid instability. This gives a difference in $\eta$ of up to three orders of magnitude. In \citet{2013PhPl...20f1202L} the diffusivity was computed for chromospheric temperatures and of order $3\times10^3$~m$^2$\,s$^{-1}$ before the onset of the plasmoid instability, much closer to our values. The reconnection rate then reached $\approx0.05$, only a factor $\sim2$ above our value. Thus, the high reconnection rate found in \citet{2012ApJ...760..109L} results in part from the high diffusivity chosen in this paper. The reconnection rate of $\approx0.05$ in \citet{2013PhPl...20f1202L} is due to recombination effects and higher but comparable to the reconnection rate we find during the plasmoid instability. After the onset of the plasmoid instability in \citet{2013PhPl...20f1202L}, the reconnection rate shows a steep increase, which could only be followed up to $\approx0.06$, since the current sheet width then reached the resolution limit. The steep increase can be taken as an indication that the plasmoid instability is potentially very important in the weak coupling description of chromospheric reconnection as well. Additionally we note that \citeauthor{2012ApJ...760..109L} used the magnetic field and Alfv\'en velocity in the upstream region not very far from the current sheet to normalize the reconnection rate, while we used the asymptotic inflow values. Our reconnection rates increase by a factor 2--3 if we use the magnetic field and Alfv\'en velocity at distances in the range 5--20 current sheet half widths ($(2.5\mbox{--}10)\delta x$). Since the initial plasma density at the initial main reconnection X-point ($y\approx1.5L_0$) in our models ($\sim10^{20}$m$^{-3}$ in Models~I and III) is more than 10 times higher than that in their models ($6\times10^{18}$m$^{-3}$), and the initial temperature($T_0 = 7000$~K) is lower than in their models ($T_0=8500$~K), our models represent the conditions at somewhat smaller heights in the chromosphere.

Our simulations do not include thermal conduction, which can generally be expected to be energetically less important than radiative cooling at chromospheric temperatures of $\sim10^4$~K which dominate in the major part of our simulated volumes \citep{1992A&A...253..557L}. Much higher temperatures and steep temperature gradients are built up in the plasmoids and at the slow-mode shocks and X-points. Conductive energy transport from these regions into the inflow regions and into the big plasmoid at the bottom end of the current sheet is largely directed across the magnetic field (conduction in the third direction, parallel to the field, is not relevant in our 2.5-dimensional simulations). Thermal conduction by the charged particles is very strongly suppressed in the direction perpendicular to the field; however, the contribution by the neutral component is not influenced. Deviations from local thermodynamic equilibrium (LTE) allow a neutral component to be present even at the high temperatures that we find in the plasmoids. To estimate the potential magnitude of thermal conduction for the conditions of our simulations, we follow \citet{1961ApJ...134...63O}. They find that the thermal conductivity across the field is provided by the neutral component in the whole temperature range relevant here and give the coefficient of thermal conductivity in a pure hydrogen gas as
\begin{equation}
 \kappa = \frac{75}{64}k_\mathrm{B}(\frac{k_\mathrm{B}T}{\pi m_\mathrm{i} M_\mathrm{H}})^{1/2}\frac{n_\mathrm{n}}{\sigma_\mathrm{nn}(n_\mathrm{n}+n_\mathrm{i})},
\end{equation}   
where $M_\mathrm{H}=1$ is the molecular weight and $\sigma_\mathrm{nn}=3.21\times10^{-20}$~m$^2$ is the neutral-neutral collisional cross section (note that \citeauthor{1961ApJ...134...63O} use $\sigma_\mathrm{nn}^2$ to designate the cross section). Since the ionization degree is small and does not change with time, $n_\mathrm{n}/(n_\mathrm{n}+n_\mathrm{i}) \simeq 1$ in our models. Under these conditions, the coefficient of thermal conductivity is evaluated as
\begin{equation}
 \kappa \simeq 2.59\times10^{-2} T^{1/2},
\end{equation}
where $T$ is measured in Kelvin and the dimension of $\kappa$ is J\,K$^{-1}$\,m$^{-1}$\,s$^{-1}$. The thermal conduction is given by
\begin{equation}
  F_C = \nabla \cdot (\kappa \nabla T).
\end{equation}
Our temperature gradient scale (slow-mode shock width) is typically $\sim500$~m. For the temperature as high as $T=80\,000$~K, the resulting thermal conduction $F_C \sim 2.59 \times 10^{-3}T^{3/2}\delta^{-2}$ is about $2.3$~J\,m$^{-3}$\,s$^{-1}$. This falls in the range of radiative cooling relevant in our models. Considering Model~III, which encompasses densities $n_\mathrm{H}\sim10^{19}\mbox{--}10^{21}$~m$^{-3}$ and corresponding ionization degrees $Y_\mathrm{i}\sim10^{-3}\mbox{--}10^{-2}$, and using $\alpha\simeq0.01$ as a conservative estimate, plasma at $80\,000$~K cools radiatively at rates of $0.35\mbox{--}350$~J\,m$^{-3}$\,s$^{-1}$. This suggests that neutral thermal conduction could limit the rise of the temperature in Model~I and in the upper part of the volume in Model~III, where the densities are relatively low. However, if the evolution of the ionization degree were also included, neutral thermal conduction can be expected to be negligible. At $T=80\,000$~K the ionization degree is very high, e.g., \citet{1961ApJ...134...63O} give $n_\mathrm{n}/n_\mathrm{i} \approx 3\times10^{-6}$ in their Table~5, so that the estimate of the neutral conductive heat flux drops by this factor and is thus lower that the radiative losses by at least four orders of magnitude throughout the range of parameters studied in this paper. Additionally, more recent considerations of the collisional cross sections \citep{2012ApJ...747...87K} indicate a value for $\sigma_\mathrm{nn}$ higher by an order of magnitude than given by \citeauthor{1961ApJ...134...63O}. The above estimates should be checked in future simulations of chromospheric reconnection when non-equilibrium ionization is also taken into account.

Effects of potential relevance but not yet included in our model are radiative transport and non-LTE effects, non-equilibrium ionization effects, and at very small scales the Hall effect. Additionally, the complex topology of the chromospheric magnetic field has not yet been addressed. It is currently impossible to include all of them in a single model. Perhaps the most relevant steps beyond our approximations are the inclusion of recombination effects in a two-fluid description, more complex magnetic topologies guided by the observations, and radiative transport. Regarding the latter extension, we note that several estimates of the reconnection rate in plasmas with radiative cooling in \citet{2011PhPl...18d2105U} were found to not depend sensitively on the specific properties of the cooling process. Also, the effect of radiative losses is only a weak to moderate one in the computations presented here. These facts may be taken as an indication that the inclusion of radiative transport in the optically thick case would not strongly influence most of our results.

\section*{ACKNOWLEDGMENTS}
We would like to thank Kai Germaschewski for his help in analyzing and displaying multi-level AMR data and James Leake, Mark Linton, Udo Ziegler, and an anonymous referee for constructive, very helpful comments.  This research is supported by NSFY (Grant No. 11203069), the Yunnan Province (Grant No.  2011FB113), the key Laboratory of Solar Activity grant KLSA201404, China Scholarship Council (CSC) No. 201404910269, Western Light of Chinese Academy of Sciences, Program 973 grants 2011CB811403 and 2013CBA01503, NSFC grant 11273055, NSFC grant 11333007, CAS grant KJCX2-EW-T07 and CAS grant XDB09000000.  B.K. acknowledges the hospitality of the solar group at the Yunnan Observatories, where most of his work was carried out, and the associated support by the Chinese Academy of Sciences under grant no.~2012T1J0017. He also acknowledges support by the DFG. We have used the NIRVANA code v3.6 developed by Udo Ziegler at the Leibniz-Institut f\"ur Astrophysik Potsdam. The authors gratefully acknowledge the computing time granted by the Yunnan Observatories, and provided on the facilities at the Yunnan Observatories Supercomputing Platform, as well as the help from all faculties of the Platform.

\clearpage

\begin{figure*}
  \centerline{\includegraphics[width=0.80\textwidth, clip=]{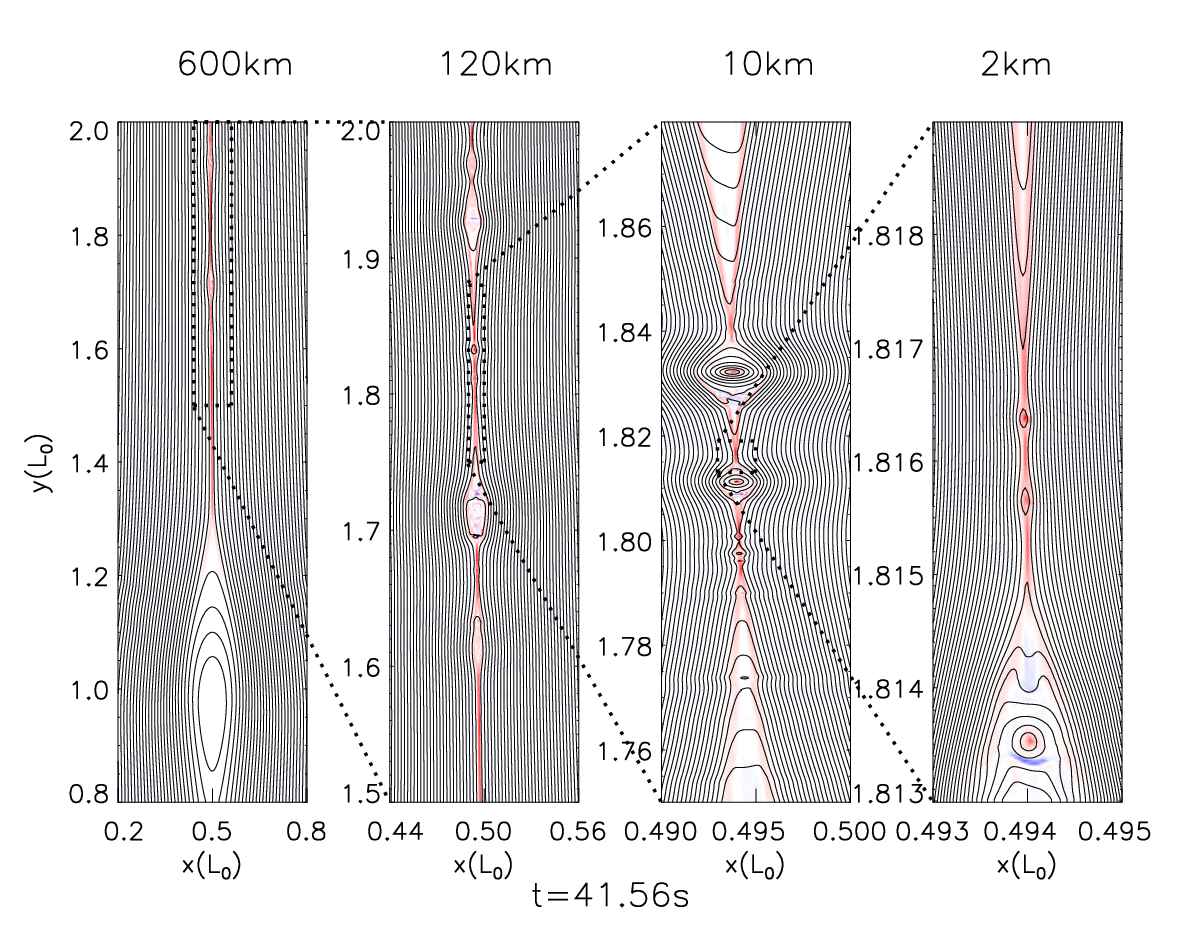}}
     \caption{Field lines and current density showing four levels of plasmoids and fragmentary current sheet sections during the cascade to small scales for Model~III, Case~I at   $t=41.56$~s.}
   \label{fig.2}
\end{figure*}

\begin{figure*}
  \centerline{\includegraphics[width=0.40\textwidth, clip=]{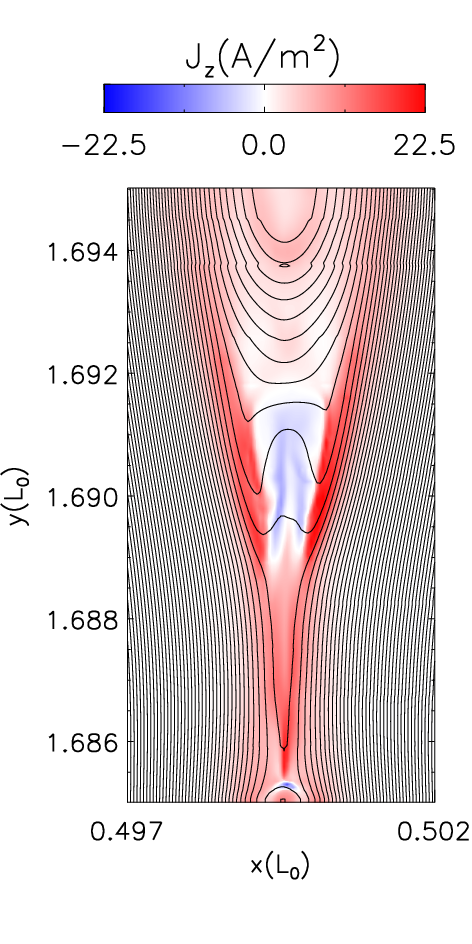}
                        \includegraphics[width=0.40\textwidth, clip=]{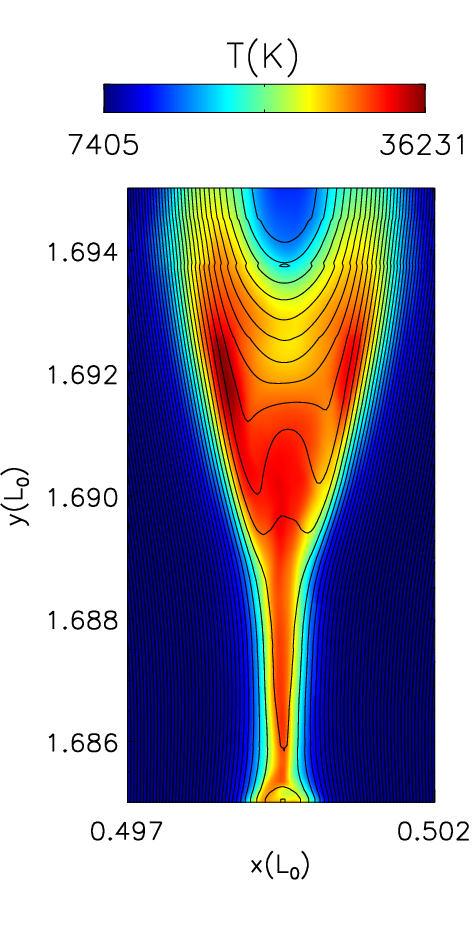}}
  \caption{\textit{Left:} Field lines and current density of a plasmoid in Model~I, Case~B at $t=31.3$~s.
          \textit{Right:} Heating at the slow-mode shocks at the edge of the plasmoid.}
 \label{fig.3}
\end{figure*}

\begin{figure*}
   \centerline{\includegraphics[width=0.6\textwidth, clip=]{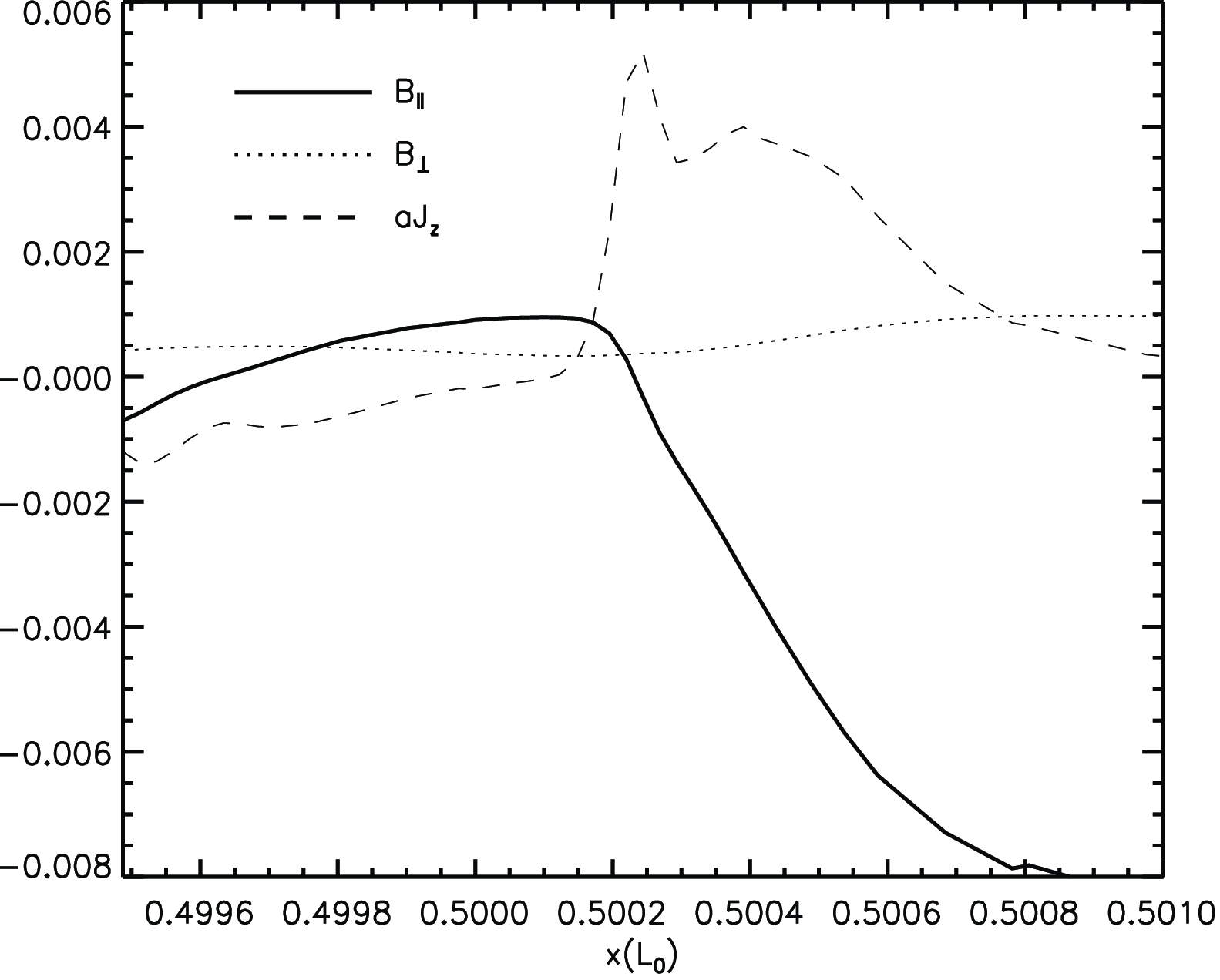}}
   \caption{Magnetic field components parallel and perpendicular to the slow-mode shock and current density along a cut line at $y=1.691~L_0$ in Fig.~\ref{fig.3} (with a scale factor $a=0.00025$ adjusting it to the range of the plot). }
  \label{fig.4}
\end{figure*}

\begin{figure*}
   \centerline{ \includegraphics[width=0.4\textwidth, clip=]{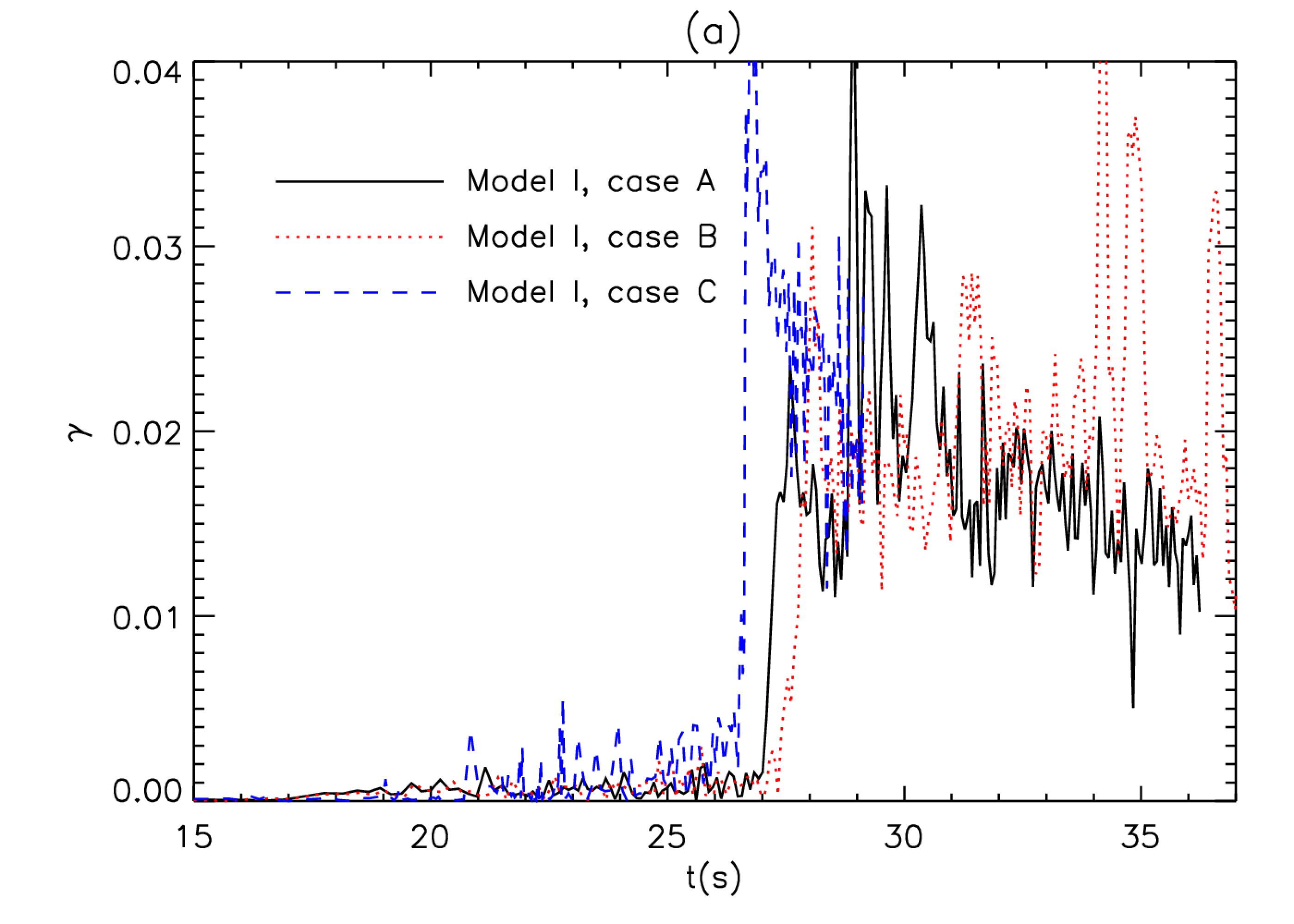}
                      \includegraphics[width=0.4\textwidth, clip=]{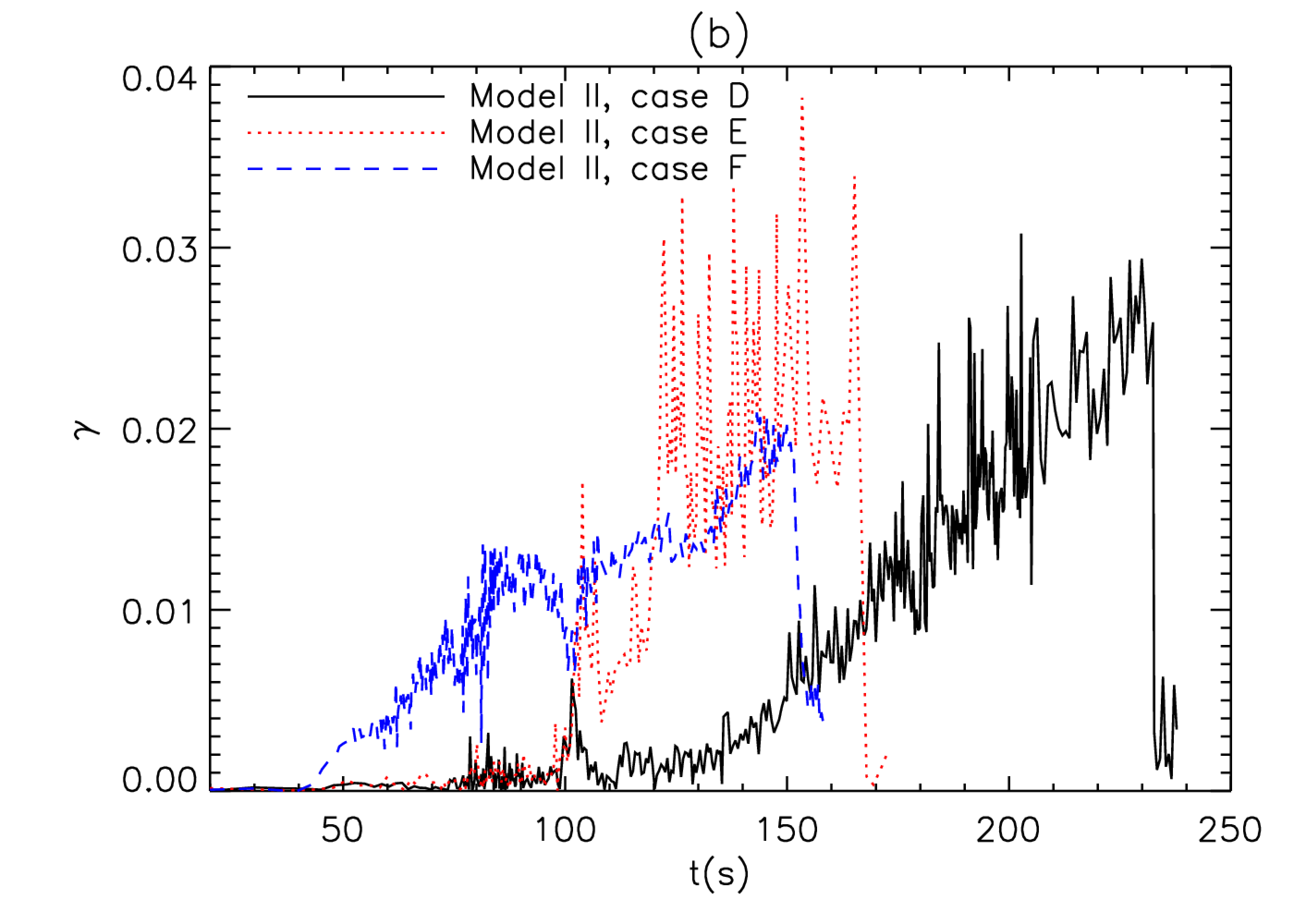}
                        \includegraphics[width=0.4\textwidth, clip=]{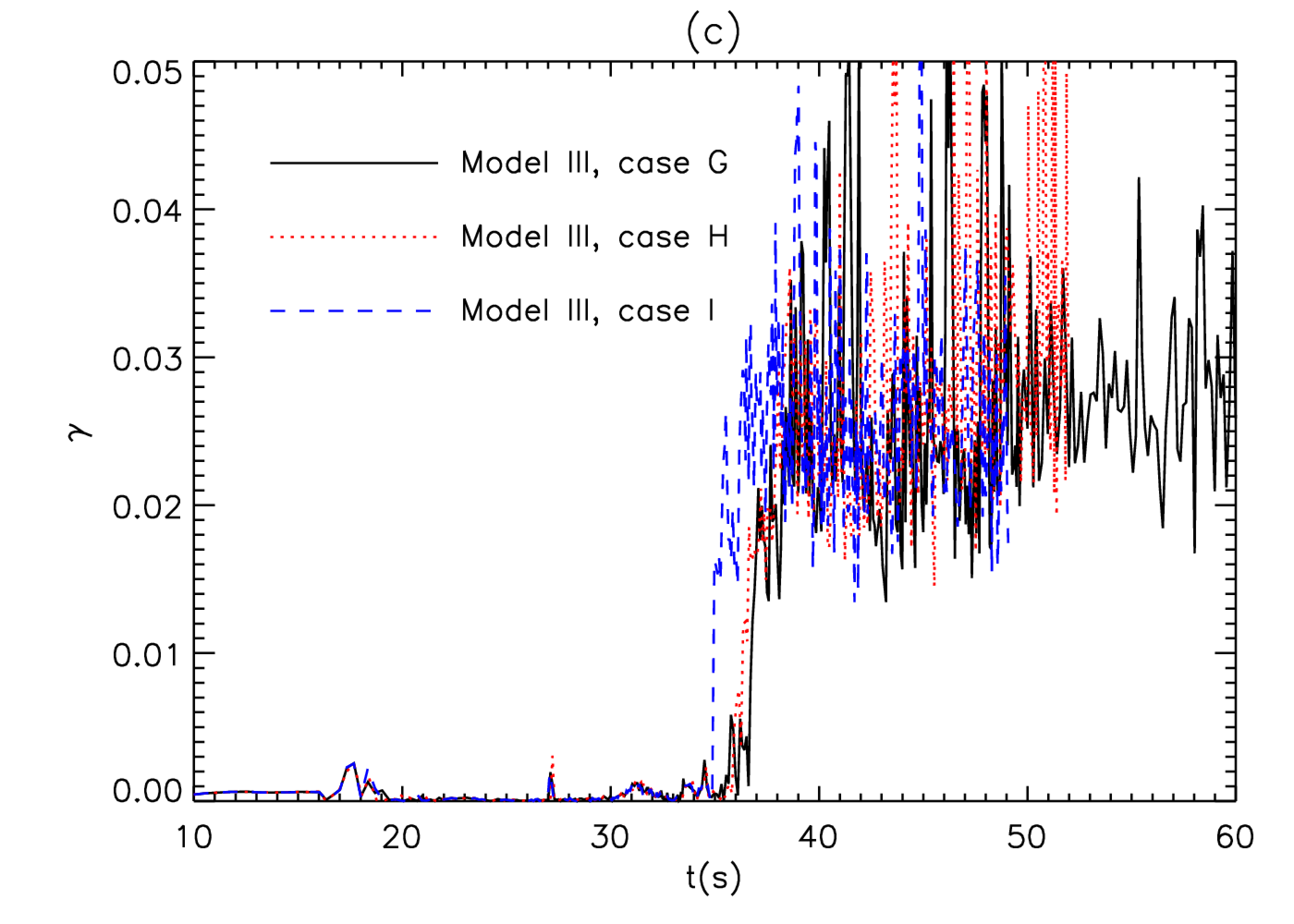}}
    \caption{Normalized reconnection rate in (a) Model~I with strong guide field and without gravity, (b) Model~II without guide field and without gravity, and (c) Model~III with strong guide field and gravity.}
 \label{fig.5}
\end{figure*}
 
\begin{figure*}
 \centerline{ \includegraphics[width=0.4\textwidth, clip=]{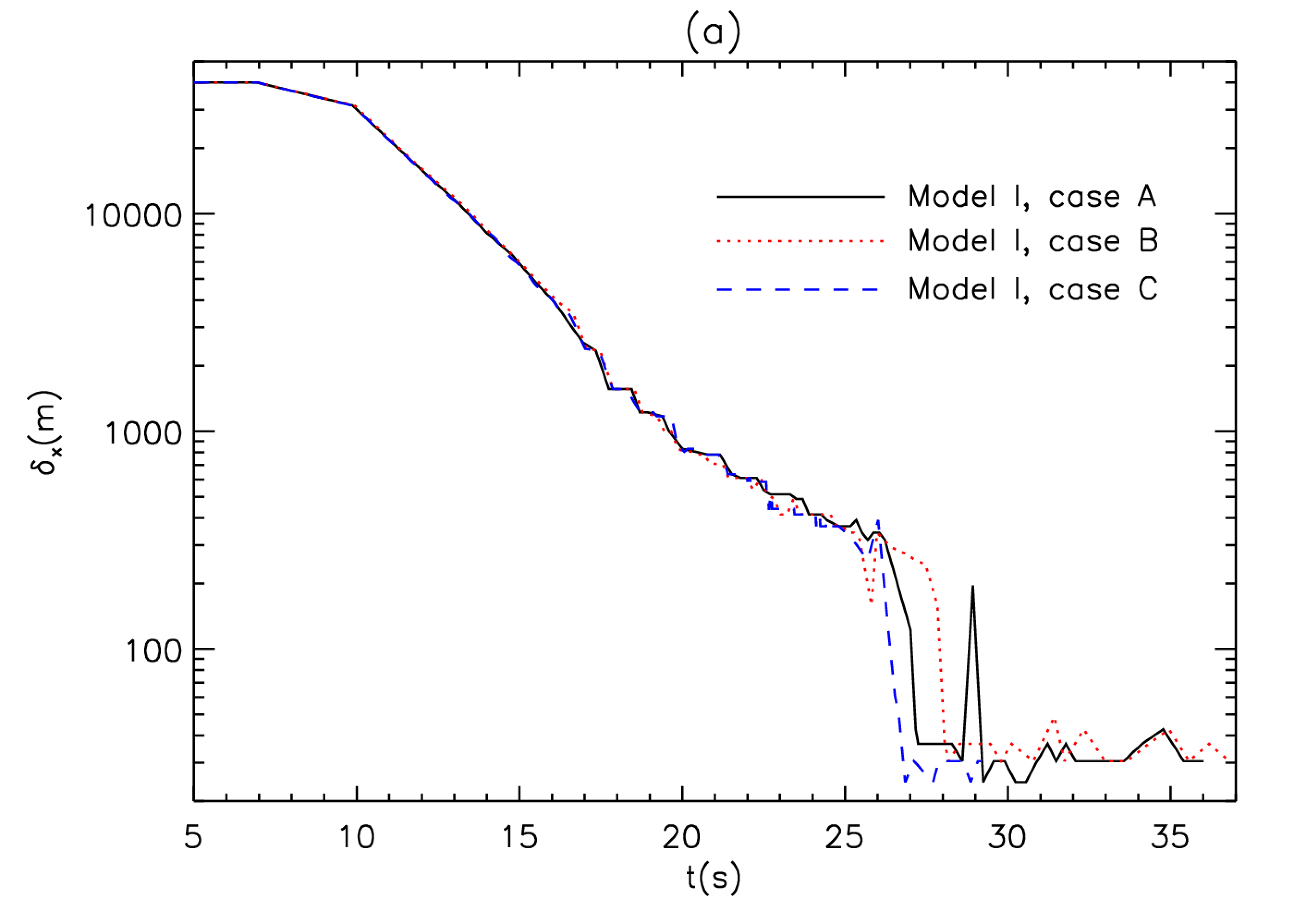}
                     \includegraphics[width=0.4\textwidth, clip=]{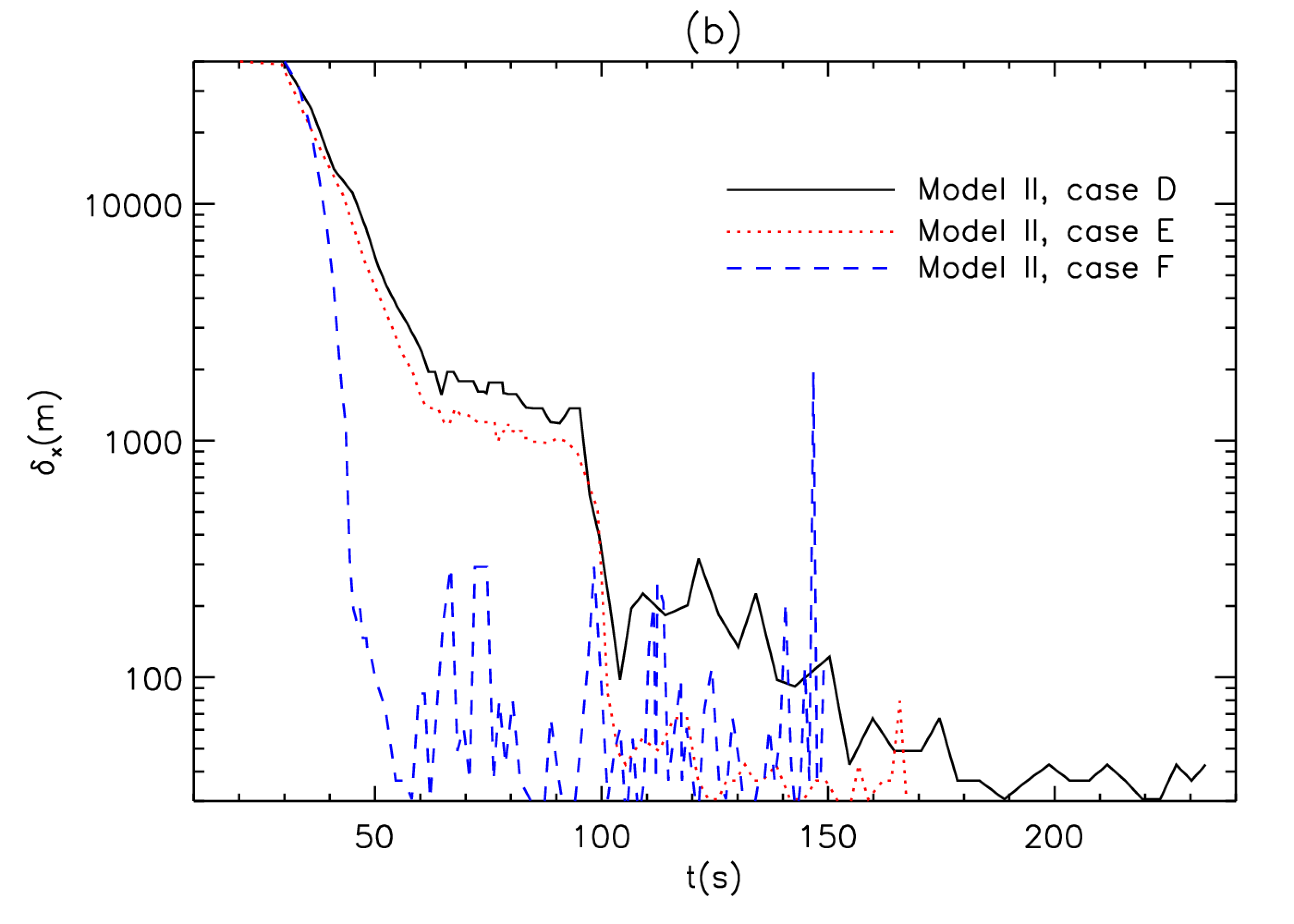}
                     \includegraphics[width=0.4\textwidth, clip=]{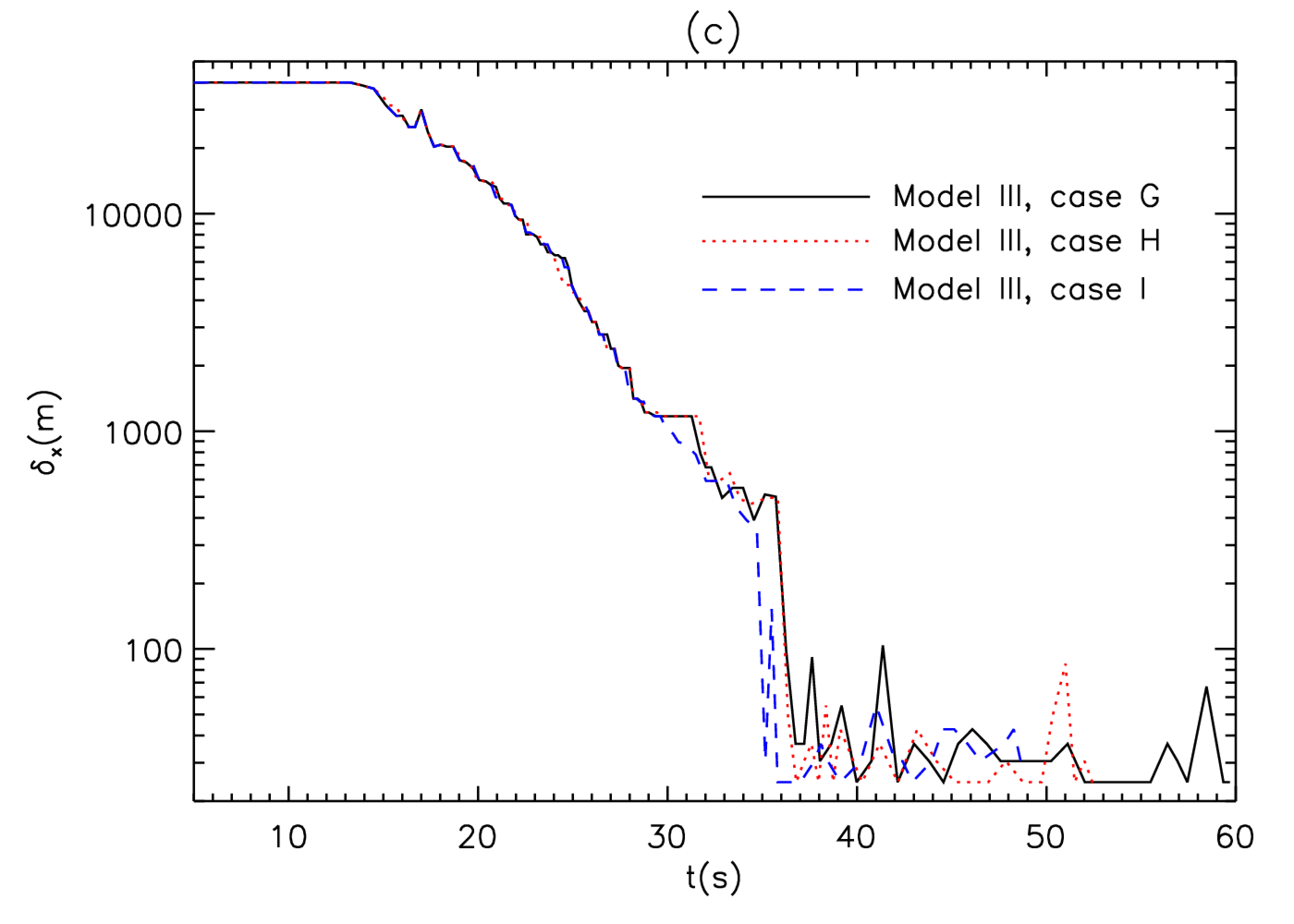}}
  \caption{Current sheet width (FWHM) at the main $X$-point in (a) Model~I, (b) Model~II and (c) Model~III.}
 \label{fig.6}
\end{figure*}
 
\begin{figure*}
   \centerline{ \includegraphics[width=0.4\textwidth, clip=]{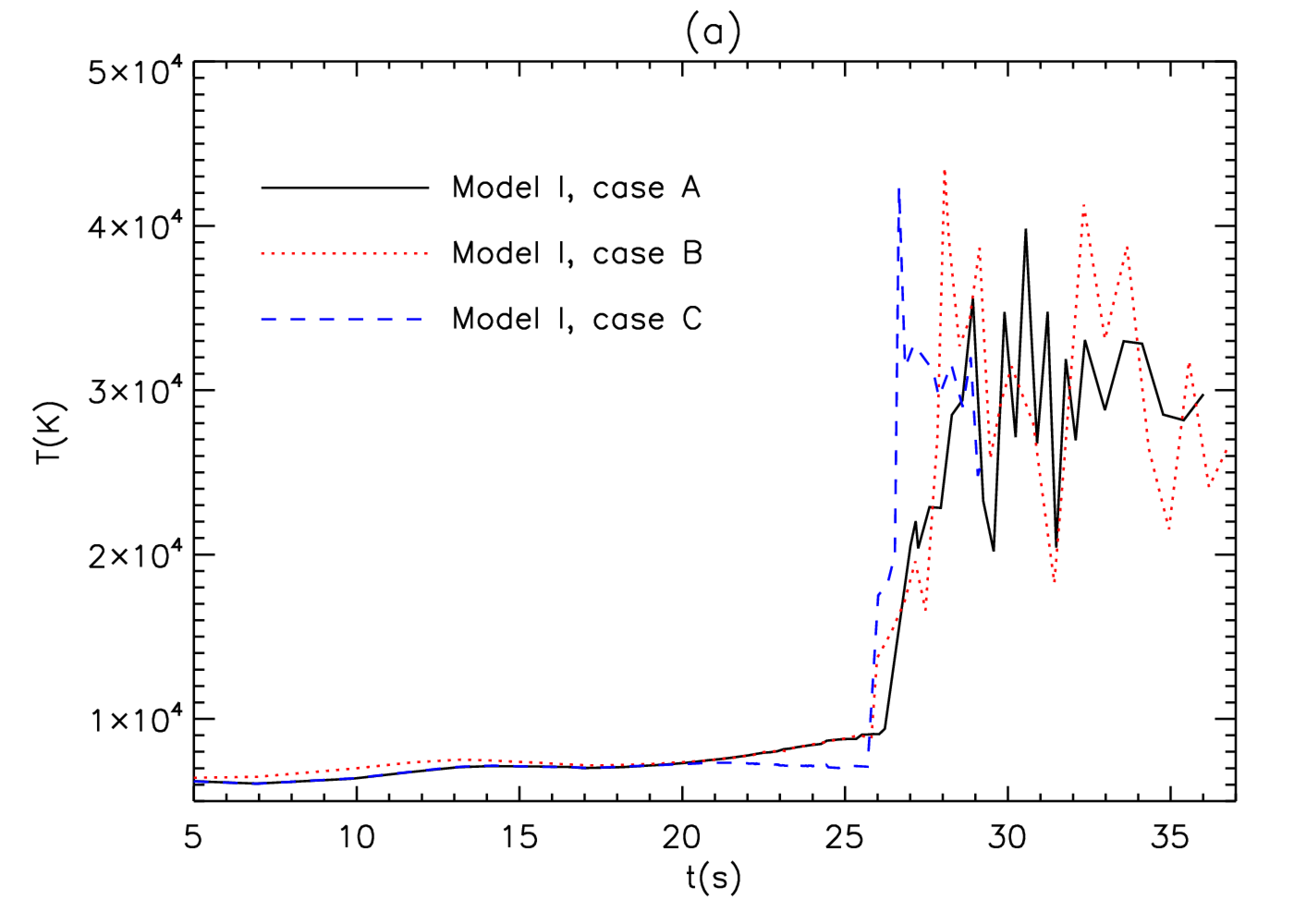}
                     \includegraphics[width=0.4\textwidth, clip=]{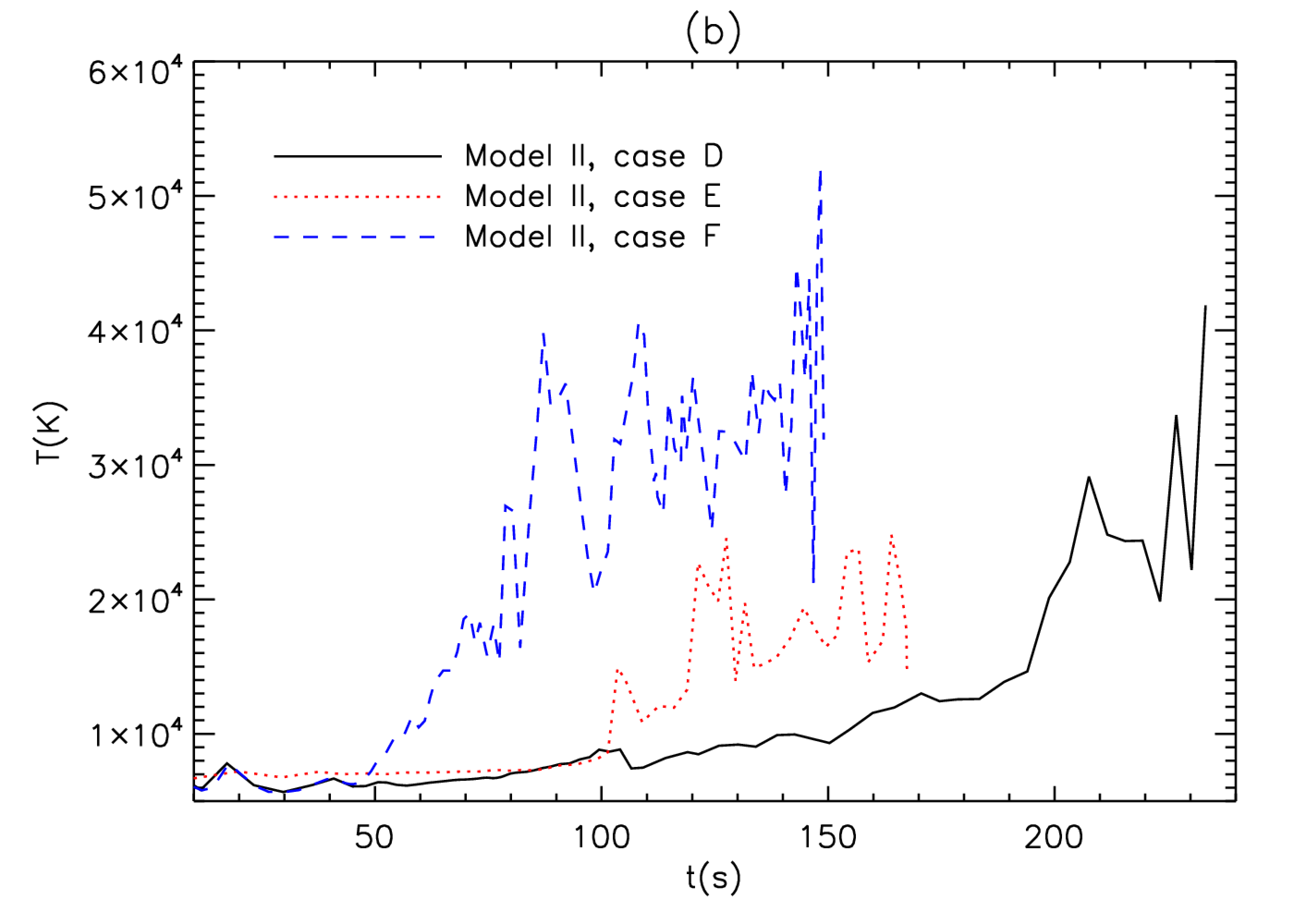}
                     \includegraphics[width=0.4\textwidth, clip=]{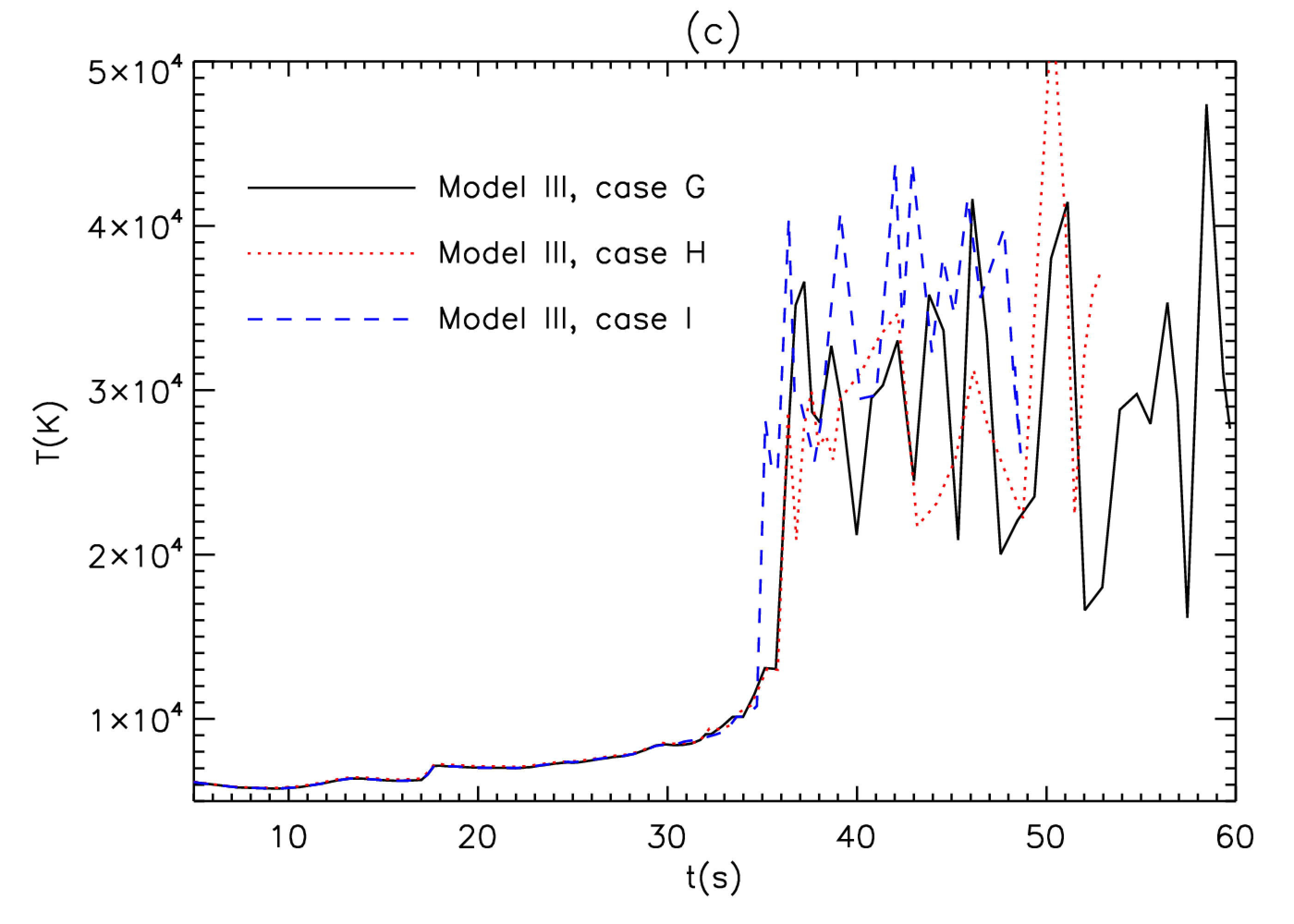}}
   \caption{Temperature at the main $X$-point in (a) Model~I, (b) Model~II and (c) Model~III.}
\label{fig.7}
\end{figure*}

 \clearpage 

\begin{figure*}
 \centerline{\includegraphics[width=0.9\textwidth, clip=]{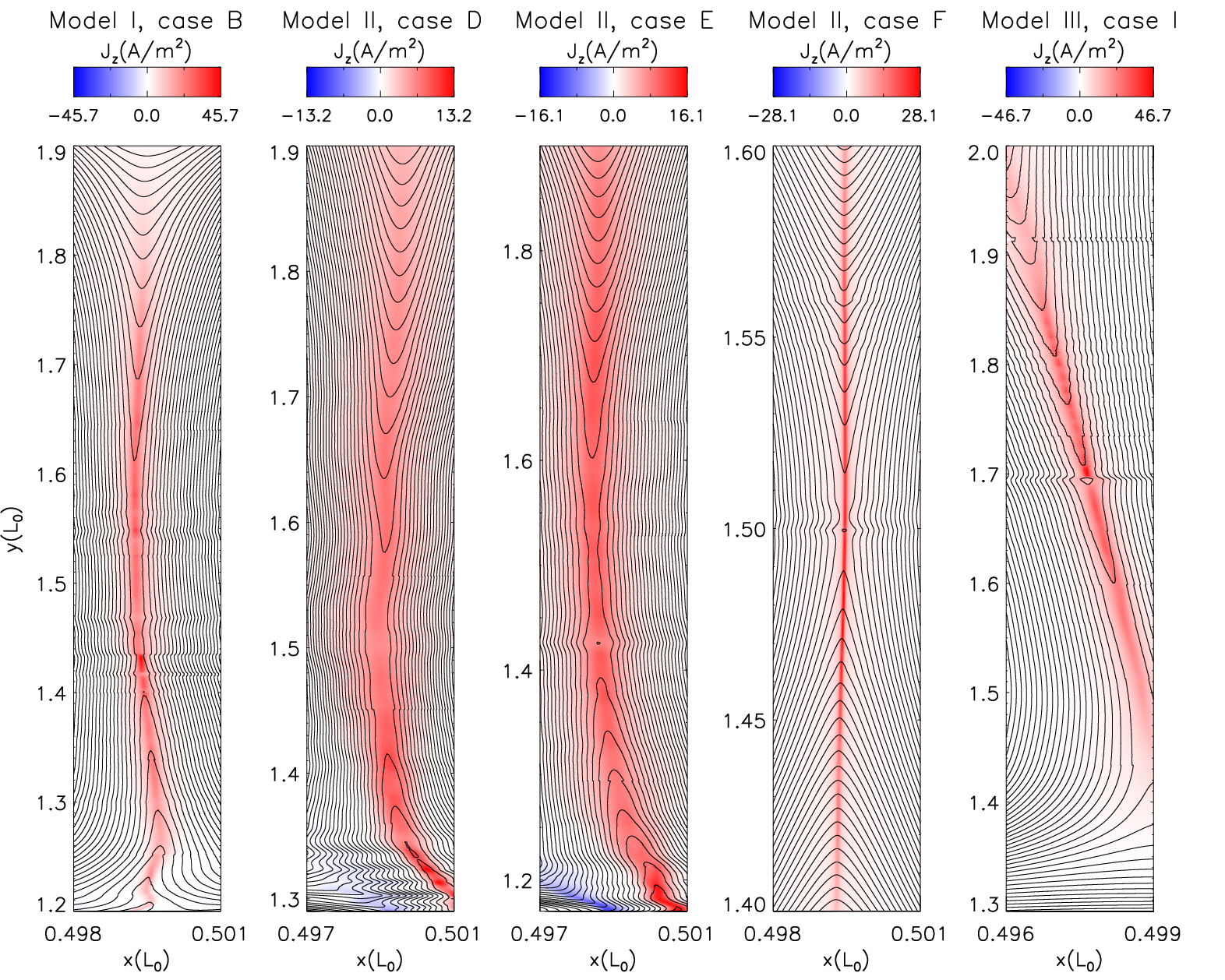}}
 \caption{Field lines and current density at the times of plasmoid instability onset in Model~I for Case~B ($t=26.4$~s), Model~II for Cases~D--F ($t=92.5$~s,  $84.0$~s and $50.5$~s, respectively), and Model~III for Case~I ($t=33.4$~s), indicating the aspect ratio at the onset of the plasmoid instability in each case.} 
  \label{fig.8}
\end{figure*}

\begin{figure*}
    \centerline{\includegraphics[width=0.35\textwidth, clip=]{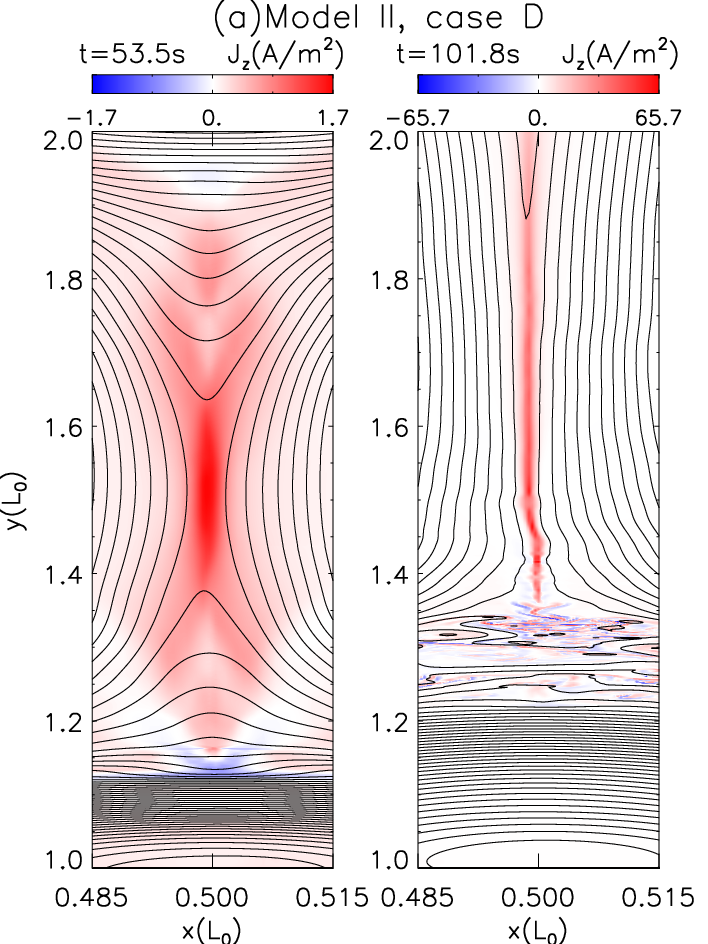}
                       \includegraphics[width=0.35\textwidth, clip=]{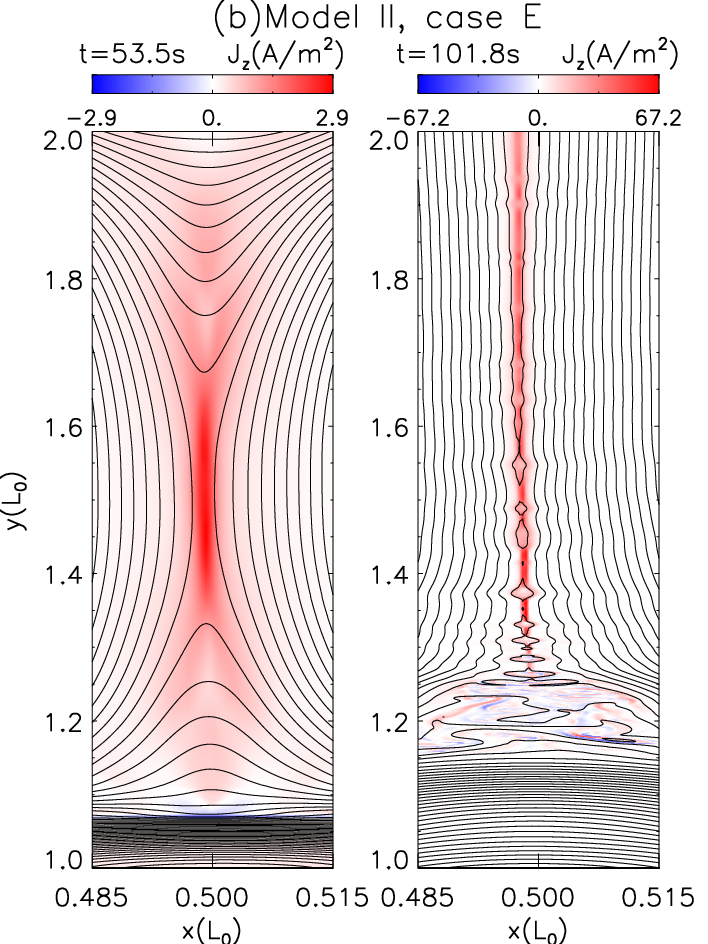}
                       \includegraphics[width=0.35\textwidth, clip=]{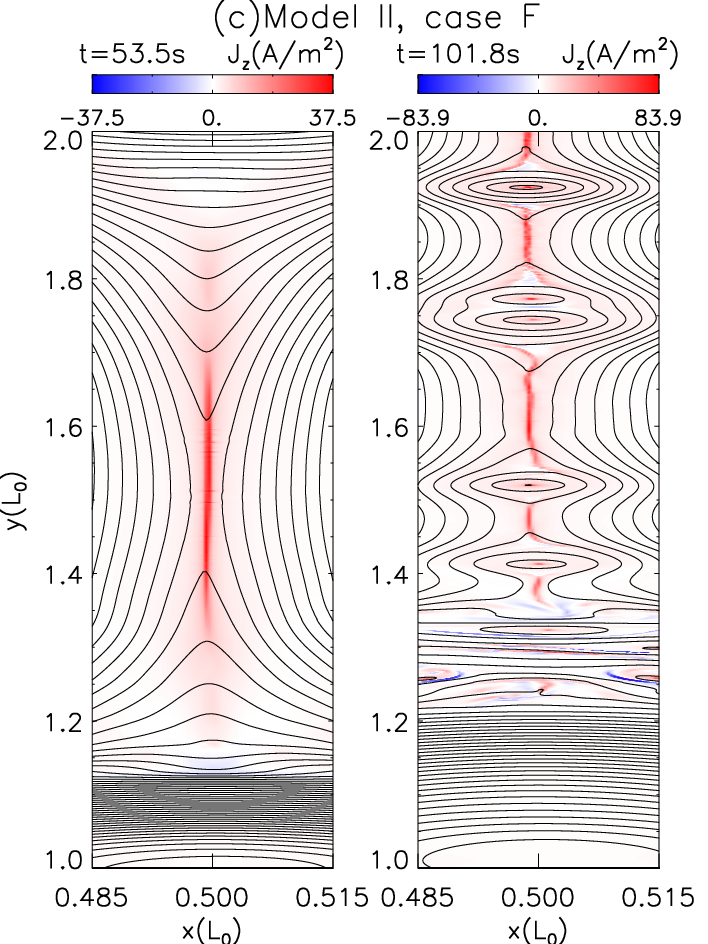}}
   \centerline{\includegraphics[width=0.35\textwidth, clip=]{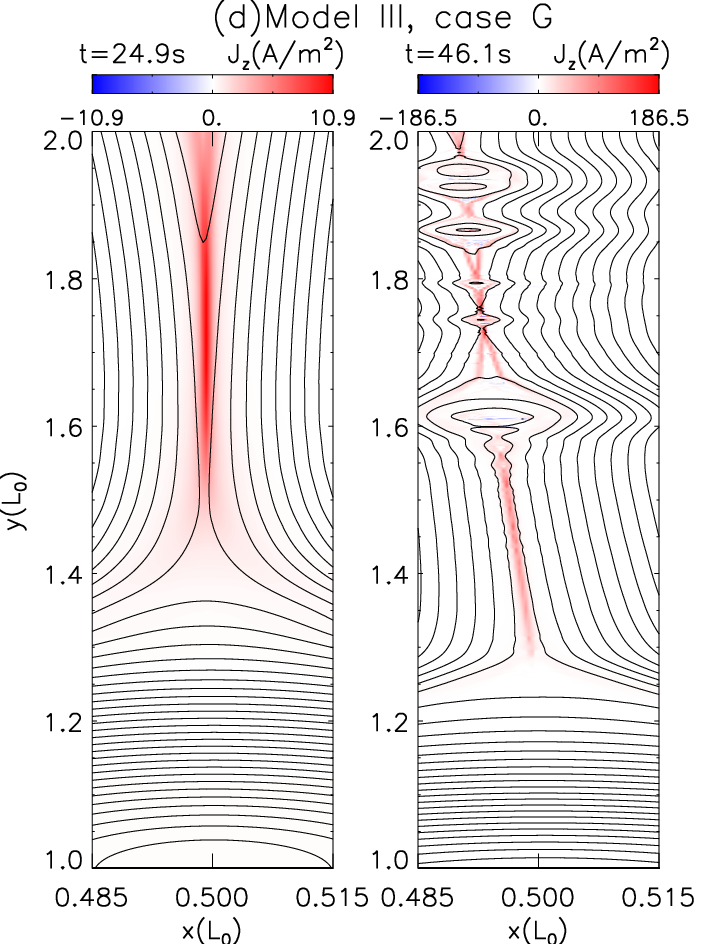}
                       \includegraphics[width=0.35\textwidth, clip=]{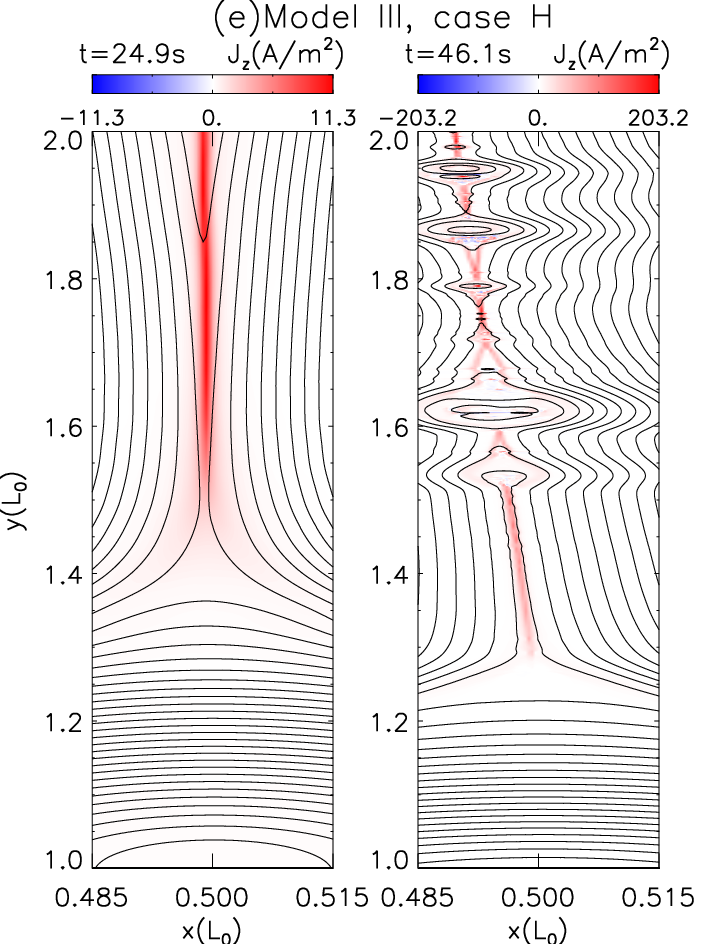}
                       \includegraphics[width=0.35\textwidth, clip=]{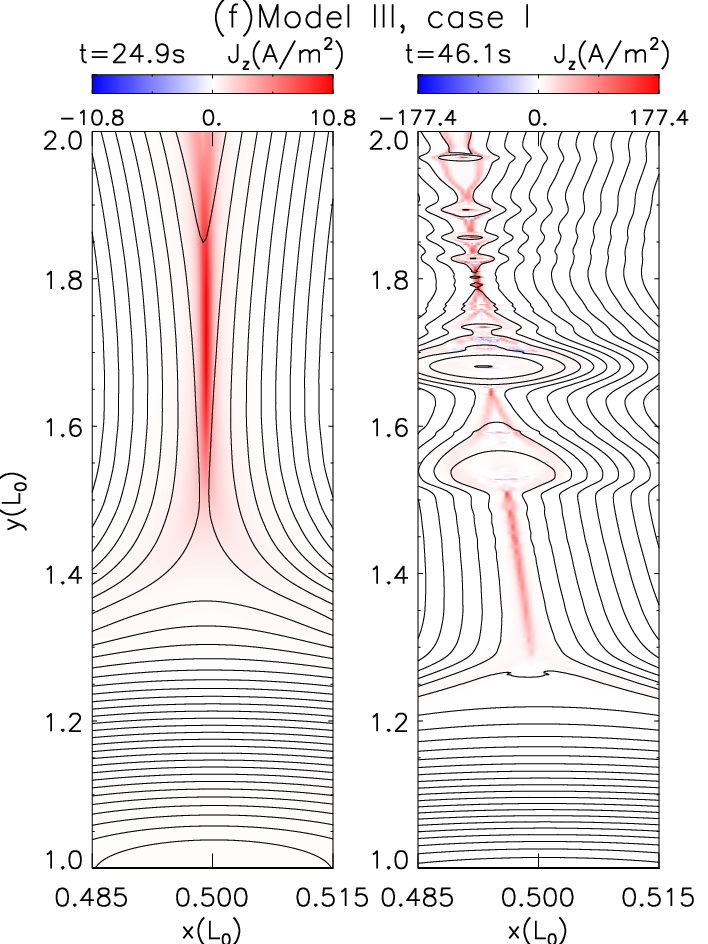}}
               \caption{Field lines and current density at two times in Model~II for Cases~D--F, and Model~III for Cases~G--I. The assignment of color to current density is adapted in each panel. The display is expanded in $x$ direction to show the plasmoids clearly.}
   \label{fig.9}
\end{figure*}

\begin{figure*}
    \centerline{\includegraphics[width=0.35\textwidth, clip=]{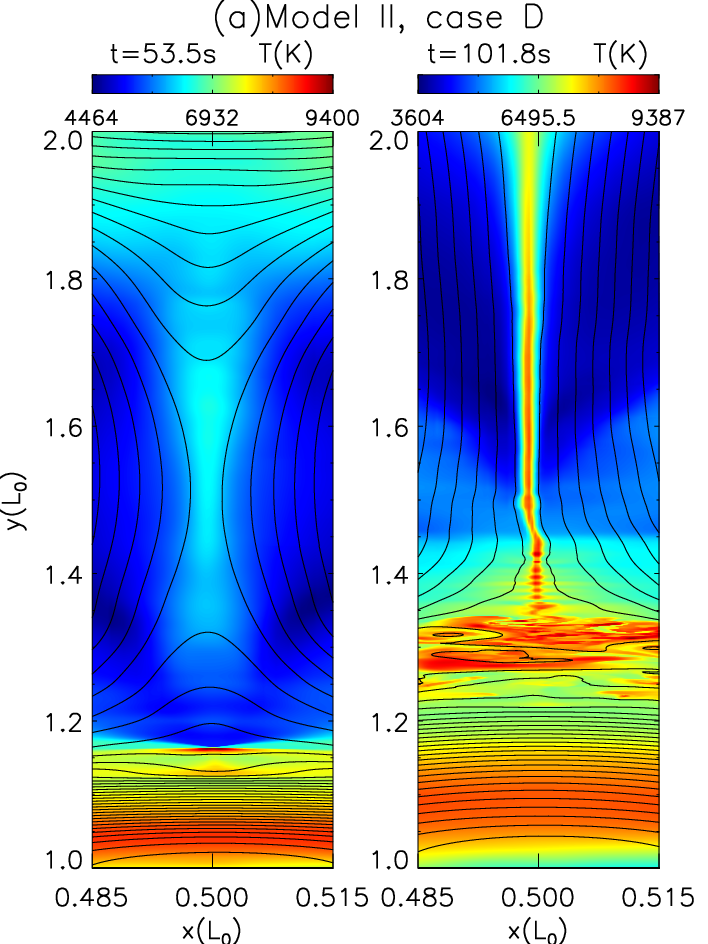}
                        \includegraphics[width=0.35\textwidth, clip=]{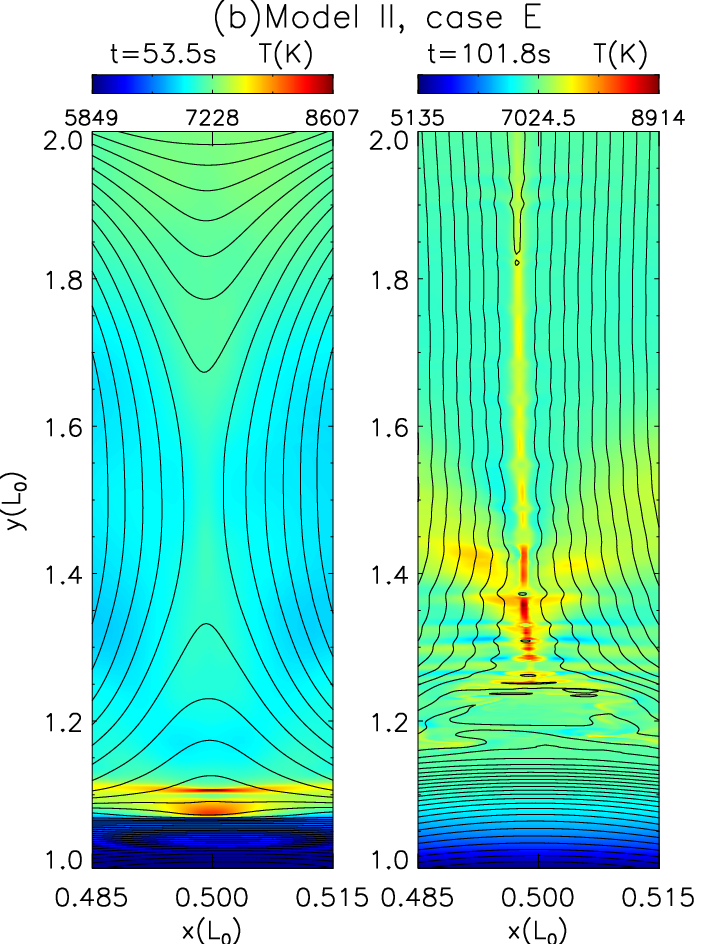}
                       \includegraphics[width=0.35\textwidth, clip=]{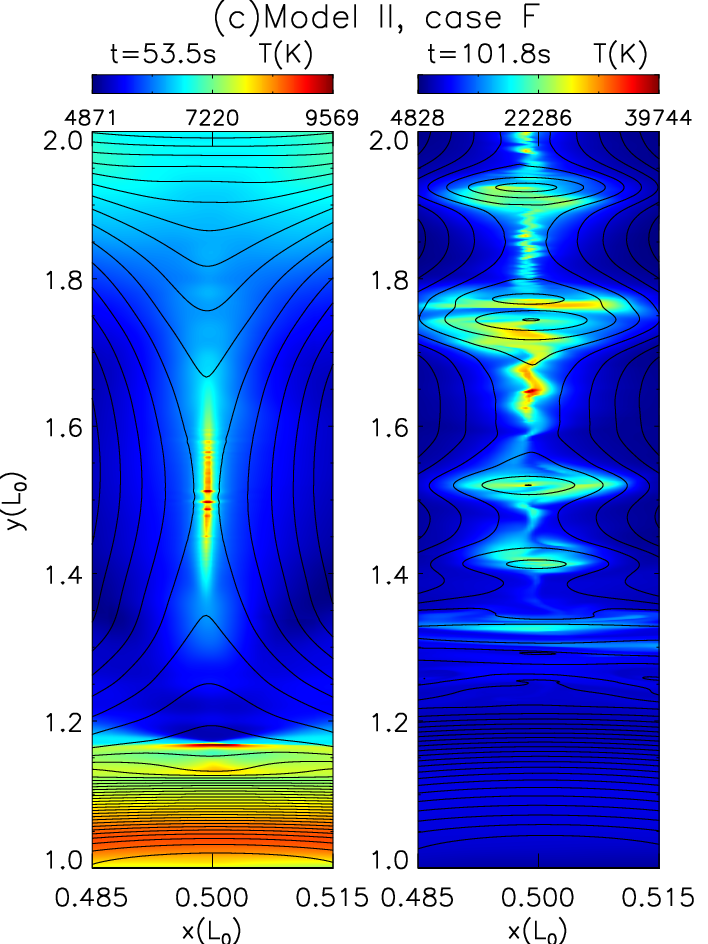}}
   \centerline{\includegraphics[width=0.35\textwidth, clip=]{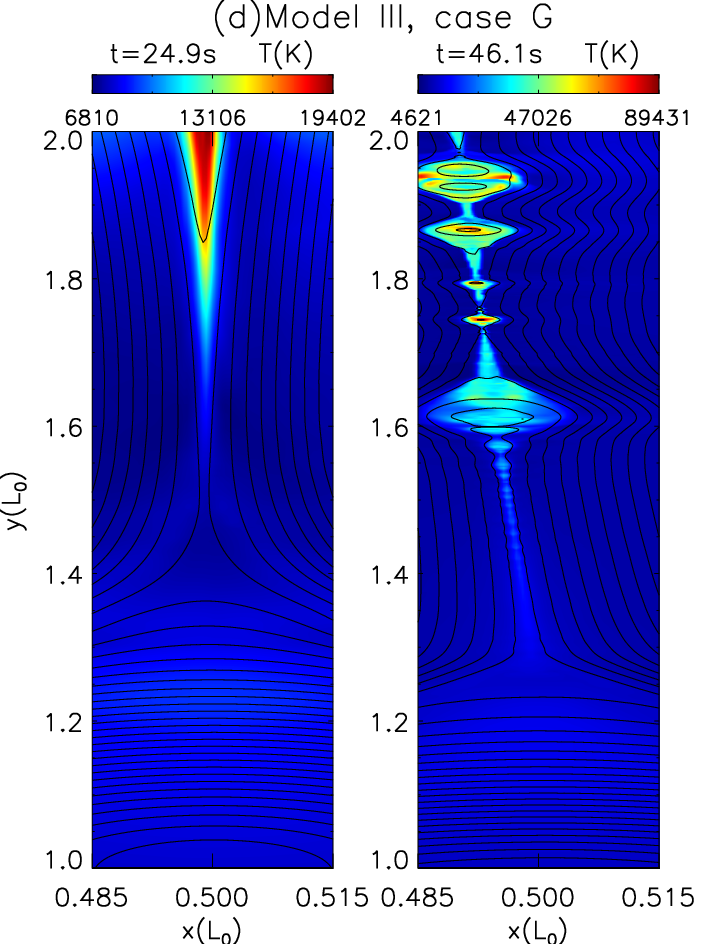}
                       \includegraphics[width=0.35\textwidth, clip=]{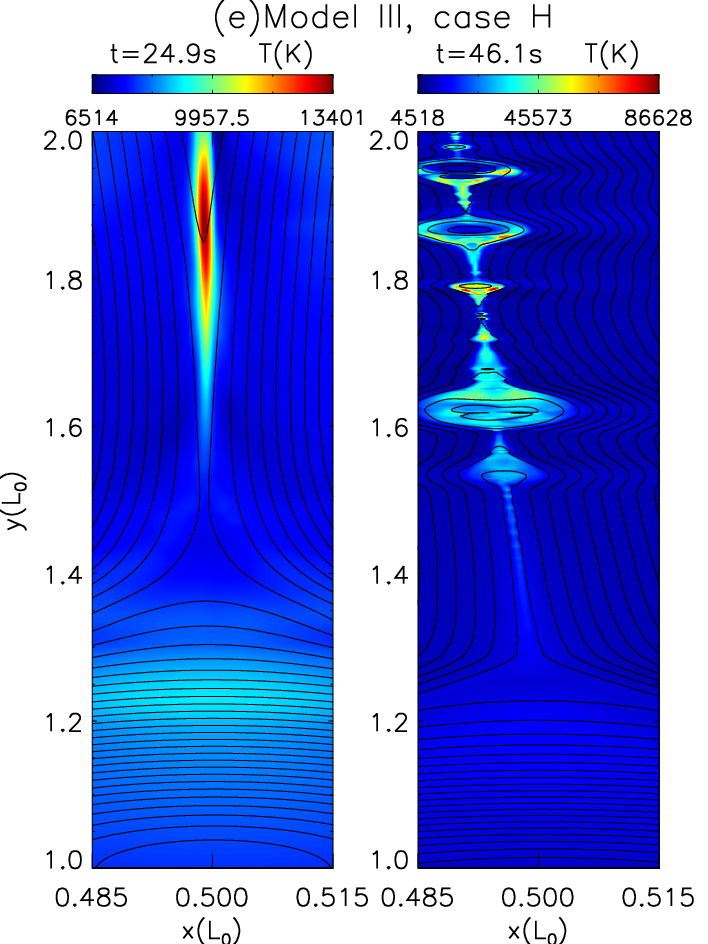}
                       \includegraphics[width=0.35\textwidth, clip=]{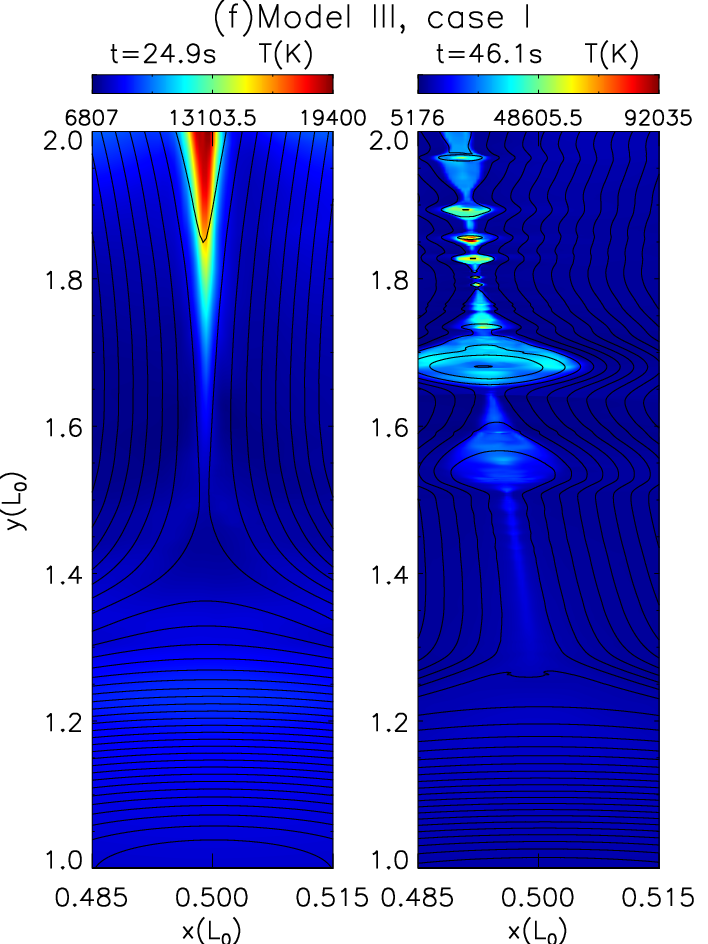}}                    
  \caption{Field lines and temperature for the same cases and times as in Fig.~\ref{fig.9}. The assignment of color to temperature is also adapted in each panel.}
 \label{fig.10}
\end{figure*}

\begin{figure*}
   \centerline{ \includegraphics[width=0.4\textwidth, clip=]{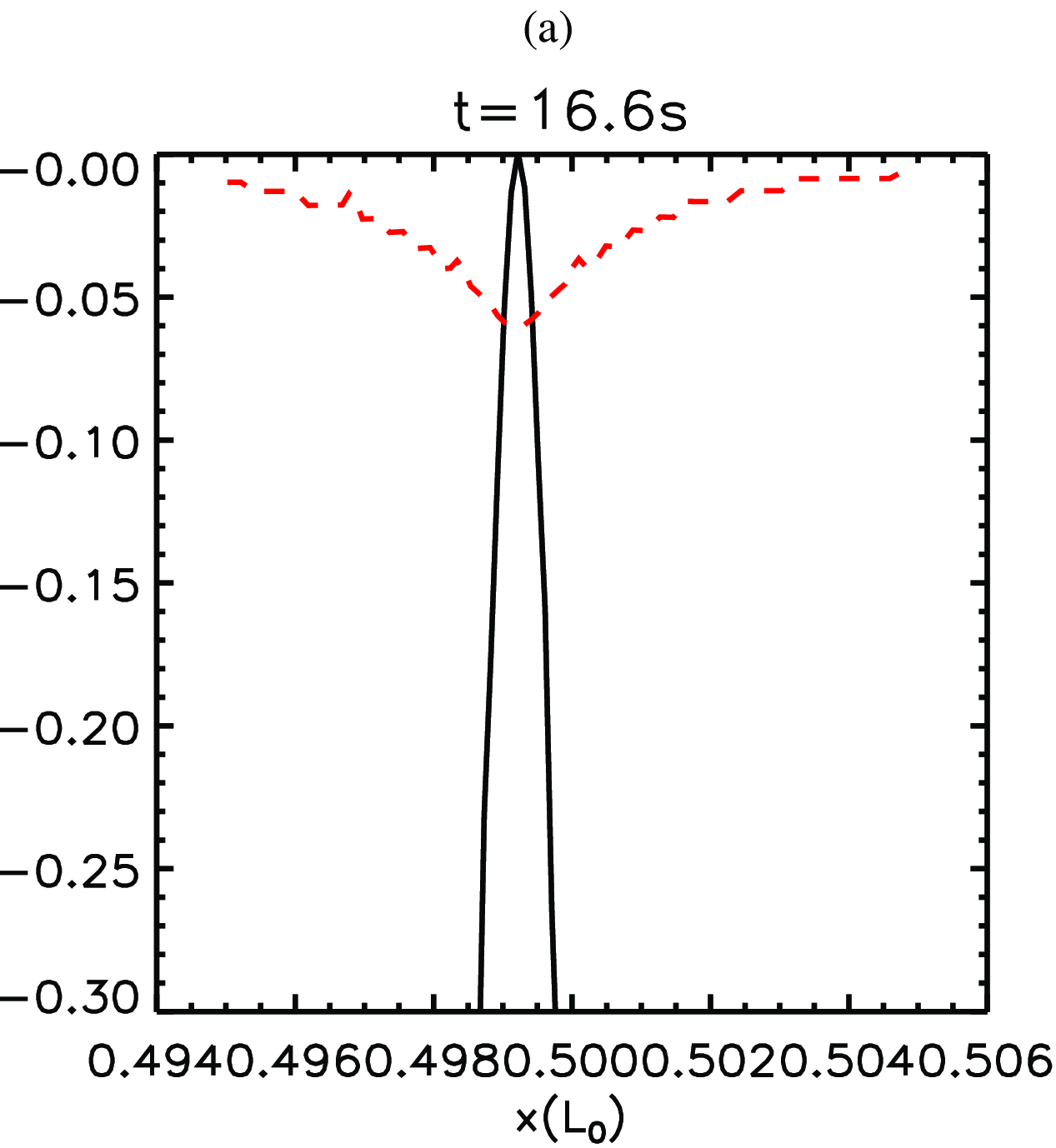}
                        \includegraphics[width=0.4\textwidth, clip=]{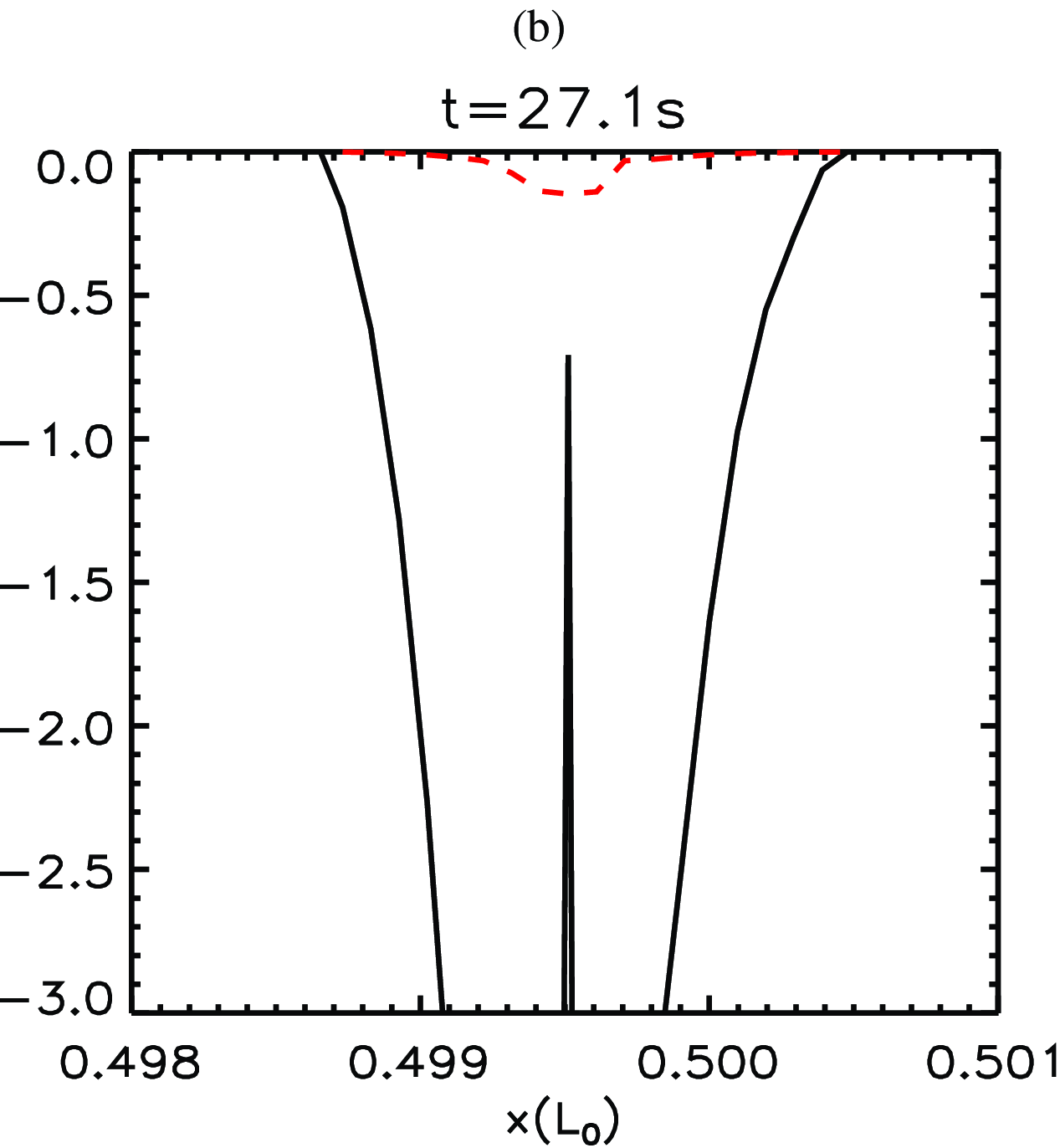}}
   \centerline{\includegraphics[width=0.5\textwidth, clip=]{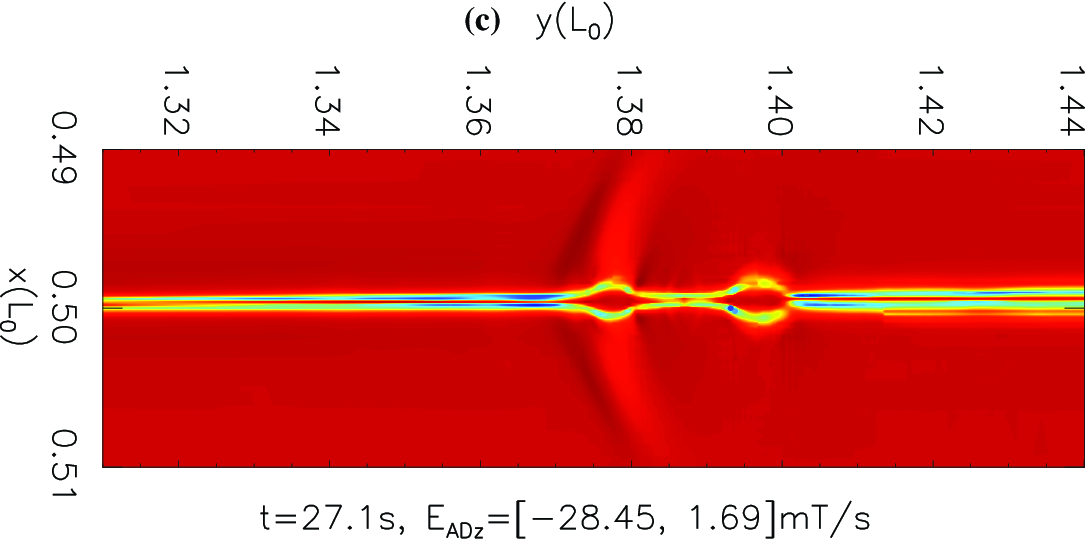}}
   \caption{ Profiles in $x$ direction of the $z$ component of the ambipolar diffusion field $E_{ADz}$ (solid black line) and $-(\eta \nabla \times \textbf{B})_z$ (red dash line) for Model~I, Case~C at the main X-point at (a) $t=16.6$~s and (b) $t=27.1$~s. (c) Spatial distribution of $E_{ADz}$ around the main X-point at $t=27.1$~s.}
 \label{fig.11}
\end{figure*}

\begin{figure*}
   \centerline{\includegraphics[width=0.6\textwidth, clip=]{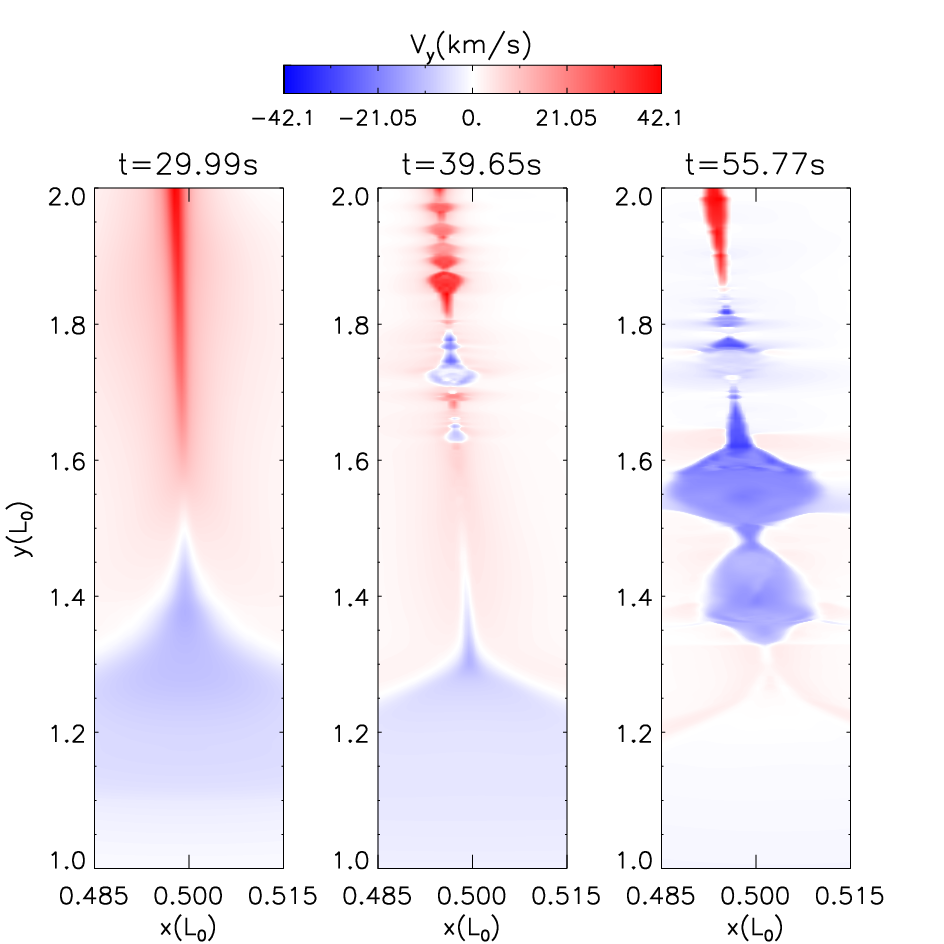}}
   \caption{Vertical velocity component for Model~III, Case~H.}
    \label{fig.12}
\end{figure*}

---------------------------


\begin{thebibliography}{99}

 \bibitem[\protect\citeauthoryear{B{\'a}rta et al.}{2011}]{2011ApJ...737...24B} B{\'a}rta, M., B{\"u}chner, J., Karlick{\'y}, M., \& Sk{\'a}la, J.\ 2011, ApJ, 737, 24
 \bibitem[\protect\citeauthoryear{Bharti, Hirzberger, \& Solanki}{2013}]{2013A&A...552L...1B} Bharti L., Hirzberger J., Solanki S.~K., 2013, A\&A, 552, L1 
 \bibitem[\protect\citeauthoryear{Bhattacharjee et al.}{2009}]{2009PhPl...16k2102B} Bhattacharjee A., Huang Y.-M., Yang H., Rogers B., 2009, PhPl, 16, 112102
 \bibitem[\protect\citeauthoryear{Braginskii}{1965}]{1965RvPP....1..205B} Braginskii S.~I., 1965, RvPP, 1, 205 
 \bibitem[\protect\citeauthoryear{Brandenburg \& Zweibel}{1994}]{1994ApJ...427L..91B} Brandenburg A., Zweibel E.~G., 1994, ApJ, 427, L91
 \bibitem[\protect\citeauthoryear{Brosius \& Holman}{2009}]{2009APJ...692..492} Brosius J.W., Holman G.D., 2009, APJ, 692,492
 \bibitem[\protect\citeauthoryear{Cameron, V{\"o}gler, \& Sch{\"u}ssler}{2007}]{2007IAUS..239..475C} Cameron R., V{\"o}gler A., Sch{\"u}ssler M., 2007, IAUS, 239, 475 
 \bibitem[\protect\citeauthoryear{Chae, Moon, \& Park}{2003}]{2003JKAS...36S..13C} Chae J., Moon Y.-J., Park S.-Y., 2003, JKAS, 36, 13
 \bibitem[\protect\citeauthoryear{Dara et al.}{1997}]{1997A&A...322..653D} Dara H.~C., Alissandrakis C.~E., Zachariadis T.~G., Georgakilas A.~A., 1997, A\&A, 322, 653 
 \bibitem[Daughton et al.(2009)]{2009PhRvL.103f5004D} Daughton, W., Roytershteyn, V., Albright, B.~J., et al.\ 2009, Physical Review Letters, 103, 065004 
 \bibitem[\protect\citeauthoryear{Denker}{1997}]{1997A&A...323..599D} Denker C., 1997, A\&A, 323, 599 
 \bibitem[de Pontieu et al.(2007)]{2007PASJ...59S.655D} de Pontieu, B., McIntosh, S., Hansteen, V.~H., et al.\ 2007, PASJ, 59, 655
 \bibitem[\protect\citeauthoryear{Dong et al.}{2012}]{2012PhRvL.108u5001D} Dong Q.-L., Wang, S.-J., Lu, Q.-M., et al., 2012, PhRvL, 108, 215001 
 \bibitem[\protect\citeauthoryear{Dorman \& Kulsrud}{1995}]{1995ApJ...449..777D} Dorman, V.~L., \& Kulsrud, R.~M.\ 1995, ApJ, 449, 777
 \bibitem[Draine et al.(1983)]{1983ApJ...264..485D} Draine, B.~T., Roberge, W.~G., \& Dalgarno, A.\ 1983, ApJ, 264, 485
 \bibitem[Draine(1986)]{1986MNRAS.220..133D} Draine, B.~T.\ 1986, MNRAS, 220, 133
 \bibitem[\protect\citeauthoryear{Fang et al.}{2006}]{2006ApJ...643.1325F} Fang C., Tang Y.~H., Xu Z., Ding M.~D., Chen P.~F., 2006, ApJ, 643, 1325
 \bibitem[\protect\citeauthoryear{Fang, Tang, \& Xu}{2006}]{2006ChJAA...6..597F} Fang C., Tang Y.-H., Xu Z., 2006, ChJAA, 6, 597 
\bibitem[Fang et al.(2003)]{2003ASPC..289..425F} Fang, C., Chen, P.~F., \& Ding, M.~D.\ 2003, The Proceedings of the IAU 8th Asian-Pacific Regional Meeting, Volume 1, 289, 425 
 \bibitem[Finn \& Kaw(1977)]{1977PhFl...20...72F} Finn, J.~M., \& Kaw, P.~K.\ 1977, Physics of Fluids, 20, 72
 \bibitem[\protect\citeauthoryear{Forbes}{1988}]{1988SoPh..117...97F} Forbes T.~G., 1988, SoPh, 117, 97
 \bibitem[\protect\citeauthoryear{Gan \& Fang}{1990}]{1990ApJ...358..328G} Gan W.~Q., Fang C., 1990, ApJ, 358, 328
 \bibitem[\protect\citeauthoryear{Georgoulis et al.}{2002}]{2002ApJ...575..506G} Georgoulis M.~K., Rust D.~M., Bernasconi P.~N., Schmieder B., 2002, ApJ, 575, 506
 \bibitem[\protect\citeauthoryear{Gontikakis, Winebarger, \& Patsourakos}{2013}]{2013A&A...550..A16} Gontikakis C., Winebarger A.R., Patsourakos S., 2013, A\&A, 550, A16
 \bibitem[Guo et al.(2013)]{2013ApJ...771L..14G} Guo, L.-J., Bhattacharjee, A., \& Huang, Y.-M.\ 2013, \apjl, 771, LL14
 \bibitem[\protect\citeauthoryear{Heitsch \& Zweibel}{2003a}]{2003ApJ...583..229H} Heitsch F., Zweibel E.~G., 2003a, ApJ, 583, 229 
 \bibitem[\protect\citeauthoryear{Heitsch \& Zweibel}{2003b}]{2003ApJ...590..291H} Heitsch F., Zweibel E.~G., 2003b, ApJ, 590, 291
 \bibitem[\protect\citeauthoryear{Huang \& Bhattacharjee}{2010}]{2010PhPl...17f2104H} Huang Y.-M., Bhattacharjee A., 2010, PhPl, 17, 062104
 \bibitem[\protect\citeauthoryear{Huang, Bhattacharjee, \& Sullivan}{2011}]{2011PhPl...18g2109H} Huang Y.-M., Bhattacharjee A., Sullivan B.~P., 2011, PhPl, 18, 072109
 \bibitem[\protect\citeauthoryear{Jiang, Fang, \& Chen}{2010}]{2010ApJ...710.1387J} Jiang R.~L., Fang C., Chen P.~F., 2010, ApJ, 710, 1387
 \bibitem[\protect\citeauthoryear{Khomenko, Collados, \& Felipe}{2008}]{2008SoPh..251..589K} Khomenko E., Collados M., Felipe T., 2008, SoPh, 251, 589 
 \bibitem[\protect\citeauthoryear{Khomenko \& Collados}{2012}]{2012ApJ...747...87K} Khomenko E., Collados M., 2012, ApJ, 747, 87
 \bibitem[Klimchuk(2006)]{2006SoPh..234...41K} Klimchuk, J.~A.\ 2006, Sol. Phys., 234, 41
 \bibitem[\protect\citeauthoryear{Kumar, Kumar, \& Uddin}{2011}]{2011PlPhR..37..161K} Kumar P., Kumar N., Uddin W., 2011, PlPhR, 37, 161 
 \bibitem[\protect\citeauthoryear{Leake, Lukin, \& Linton}{Leake et al.}{2013}]{2013PhPl...20f1202L} Leake J.~E., Lukin V.~S., Linton M.~G., 2013, PhPl, 20, 061202 
 \bibitem[\protect\citeauthoryear{Leake et al.}{2012}]{2012ApJ...760..109L} Leake J.~E., Lukin V.~S., Linton M.~G., Meier E.~T., 2012, ApJ, 760, 109 
  \bibitem[\protect\citeauthoryear{Lin, Cranmer, \& Farrugia}{2008}]{2008JGRA..11311107L} Lin J., Cranmer S.~R., Farrugia C.~J., 2008, JGRA, 113, 11107 
 \bibitem[\protect\citeauthoryear{Lin et al.}{1992}]{1992A&A...253..557L} Lin J., Zhang Z., Wang Z., Smartt R.~N., 1992, A\&A, 253, 557
 \bibitem[Liu(2013)]{2013MNRAS.434.1309L} Liu, R.\ 2013, MNRAS, 434, 1309
 \bibitem[\protect\citeauthoryear{Liu et al.}{2009}]{2009ApJ...707L..37L}Liu W., Berger T.~E., Title A.~M., Tarbell T.~D., 2009, ApJ, 707, L37 
 \bibitem[Loureiro et al.(2007)]{2007PhPl...14j0703L} Loureiro, N.~F., Schekochihin, A.~A., \& Cowley, S.~C.\ 2007, Physics of Plasmas, 14, 100703
 \bibitem[\protect\citeauthoryear{Malyshkin \& Zweibel}{2011}]{2011ApJ...739...72M} Malyshkin L.~M., Zweibel E.~G., 2011, ApJ, 739, 72
 \bibitem[Mandt et al.(1994)]{1994GeoRL..21...73M} Mandt, M.~E., Denton, R.~E., \& Drake, J.~F.\ 1994, Geophys. Res. Lett., 21, 73
\bibitem[Markidis et al.(2013)]{2013PhPl...20h2105M} Markidis, S., Henri, P., Lapenta, G., et al.\ 2013, Physics of Plasmas, 20, 082105 
 \bibitem[\protect\citeauthoryear{Mei et al.}{2012}]{2012MNRAS.425.2824M} Mei Z., Shen C., Wu N., Lin J., Murphy N.~A., Roussev I.~I., 2012, MNRAS, 425, 2824 
 \bibitem[\protect\citeauthoryear{Milligan et al.}{2010}]{2010ApJ...713.1292M} Milligan R.~O., McAteer R.~T.~J., Dennis B.~R., Young C.~A., 2010, ApJ, 713, 1292
 \bibitem[Moore et al.(2011)]{2011ApJ...731L..18M} Moore, R.~L., Sterling, A.~C., Cirtain, J.~W., \& Falconer, D.~A.\ 2011, ApJ, 731, L18
 \bibitem[\protect\citeauthoryear{Morton}{2012}]{2012A&A...543A...6M} Morton R.~J., 2012, A\&A, 543, A6
 \bibitem[Murphy et al.(2012)]{2012ApJ...751...56M} Murphy, N.~A., Miralles, M.~P., Pope, C.~L., et al.\ 2012, \apj, 751, 56 
 \bibitem[\protect\citeauthoryear{Ni et al.}{2010}]{2010PhPl...17e2109N} Ni L., Germaschewski K., Huang Y.-M., Sullivan B.~P., Yang H., Bhattacharjee A., 2010, PhPl, 17, 052109 
  \bibitem[\protect\citeauthoryear{Ni et al.}{2012}]{2012PhPl...19g2902N} Ni L., Ziegler U., Huang Y.-M., Lin J., Mei Z., 2012, PhPl, 19, 072902 
  \bibitem[\protect\citeauthoryear{Ni et al.}{2012}]{2012ApJ...758...20N} Ni L., Roussev I.~I., Lin J., Ziegler U., 2012, ApJ, 758, 20 
 \bibitem[Ni et al.(2013)]{2013PhPl...20f1206N} Ni, L., Lin, J., \& Murphy, N.~A.\ 2013, Physics of Plasmas, 20, 061206
  \bibitem[Nishizuka et al.(2011)]{2011ApJ...731...43N} Nishizuka, N., Nakamura, T., Kawate, T., Singh, K.~A.~P., \& Shibata, K.\ 2011, ApJ, 731, 43
  \bibitem[Orrall \& Zirker(1961)]{1961ApJ...134...63O} Orrall, F.~Q., \& Zirker, J.~B.\ 1961, \apj, 134, 63
  \bibitem[\protect\citeauthoryear{Parker}{1972}]{1972ApJ...174..499P} Parker E.~N., 1972, ApJ, 174, 499 
  \bibitem[\protect\citeauthoryear{Pneuman, Solanki, \& Stenflo}{1986}]{1986A&A...154..231P} Pneuman G.~W., Solanki S.~K., Stenflo J.~O., 1986, A\&A, 154, 231 
  \bibitem[\protect\citeauthoryear{Qiu et al.}{2000}]{2000ApJ...544L.157Q} Qiu J., Ding M.~D., Wang H., Denker C., Goode P.~R., 2000, ApJ, 544, L157 
  \bibitem[Raymond et al.(1976)]{1976ApJ...204..290R} Raymond, J.~C., Cox, D.~P., \& Smith, B.~W.\ 1976, \apj, 204, 290 

  \bibitem[\protect\citeauthoryear{Savage et al.}{2010}]{2010ApJ...722..329S} Savage S.~L., McKenzie D.~E., Reeves K.~K., Forbes T.~G., Longcope D.~W., 2010, ApJ, 722, 329 
 \bibitem[\protect\citeauthoryear{Schumacher \& Kliem}{1997}]{1997AdSpR..19.1797S} Schumacher, J., \& Kliem, B.\ 1997, Adv.\ Space Res., 19, 1797
  \bibitem[Shen et al.(2011)]{2011ApJ...737...14S} Shen, C., Lin, J., \& Murphy, N.~A.\ 2011, \apj, 737, 14 
  \bibitem[\protect\citeauthoryear{Shen et al.}{2011}]{2011ApJ...735L..43S} Shen Y., Liu Y., Su J., Ibrahim A., 2011, ApJ, 735, L43 
  \bibitem[\protect\citeauthoryear{Shen et al.}{2012}]{2012ApJ...745..164S} Shen Y., Liu Y., Su J., Deng Y., 2012, ApJ, 745, 164 
  \bibitem[\protect\citeauthoryear{Shibata et al.}{2007}]{2007Sci...318.1591S} Shibata K., et al., 2007, Sci, 318, 1591 
  \bibitem[\protect\citeauthoryear{Singh et al.}{2012}]{2012ApJ...759...33S} Singh K.~A.~P., Isobe H., Nishizuka N., Nishida K., Shibata K., 2012, ApJ, 759, 33 
  \bibitem[Spitzer(1956)]{1956pfig.book.....S} Spitzer, L.\ 1956, Physics of Fully Ionized Gases, New York: Interscience Publishers, 1956
  \bibitem[\protect\citeauthoryear{Sturrock}{1999}]{1999ApJ...521..451S} Sturrock P.~A., 1999, ApJ, 521, 451
  \bibitem[Takasao et al.(2012)]{2012ApJ...745L...6T} Takasao, S., Asai, A., Isobe, H., \& Shibata, K.\ 2012, ApJ, 745, L6
  \bibitem[\protect\citeauthoryear{Teriaca, Curdt, \& Solanki}{2008}]{2008A&A...491L...5T} Teriaca L., Curdt W., Solanki S.~K., 2008, A\&A, 491, L5 

  \bibitem[\protect\citeauthoryear{Uzdensky \& McKinney}{2011}]{2011PhPl...18d2105U} Uzdensky D.~A., McKinney J.~C., 2011, PhPl, 18, 042105 
  \bibitem[van Ballegooijen \& Martens(1989)]{1989ApJ...343..971V} van Ballegooijen, A.~A., \& Martens, P.~C.~H.\ 1989, \apj, 343, 971 

  \bibitem[\protect\citeauthoryear{Vernazza, Avrett, \& Loeser}{1981}]{1981ApJS...45..635V} Vernazza J.~E., Avrett E.~H., Loeser R., 1981, ApJS, 45, 635
  \bibitem[\protect\citeauthoryear{Xu et al.}{2011}]{2011RAA....11..225X} Xu X.-Y., Fang C., Ding M.-D., Gao D.-H., 2011, RAA, 11, 225 
  \bibitem[\protect\citeauthoryear{Zachariadis, Alissandrakis, \& Banos}{1987}]{1987SoPh..108..227Z} Zachariadis T.~G., Alissandrakis C.~E., Banos G., 1987, SoPh, 108, 227
  \bibitem[\protect\citeauthoryear{Ziegler}{2011}]{2011JCoPh.230.1035Z} Ziegler U., 2011, JCoPh, 230, 1035
  \bibitem[\protect\citeauthoryear{Zwaan}{1987}]{1987ARA&A..25...83Z} Zwaan C., 1987, ARA\&A, 25, 83

  \bibitem[\protect\citeauthoryear{Zweibel et al.}{2011}]{2011PhPl...18k1211Z} Zweibel E.~G., Lawrence E., Yoo J., Ji H., Yamada M., Malyshkin L.~M., 2011, PhPl, 18, 111211 
  \bibitem[Zweibel(1989)]{1989ApJ...340..550Z} Zweibel, E.~G.\ 1989, \apj, 340, 550 

\end{thebibliography}
\end{document}